\documentclass[aps,floatfix,superscriptaddress,showpacs,nofootinbib]{revtex4}
\usepackage{graphicx}
\usepackage{amsmath}
\usepackage{amssymb}
\usepackage{bm}

\usepackage{amsmath}
\usepackage{mathrsfs}
\usepackage{curves}
\usepackage{amsthm}
\usepackage{subfigure}
\usepackage{picinpar}
\usepackage{picins}
\usepackage{amscd}

\usepackage{graphicx,color}
\usepackage{caption2}

\begin{document}


\title{Pulse vaccination in the periodic infection rate SIR epidemic model}
\author{Zhen Jin}%
\email{Corresponding author. jinzhn@263.com} \affiliation{Department
of mathematics, North University of China,\\ Taiyuan 030051, P.
R.China.}

\author{Mainul Haque}%
\email{mainul.haque@rediffmail.com} \affiliation{Department of
Mathematics, National University of Singapore, Singapore-117543.}

\author{Quanxing Liu}%
\email{liuqx315@sina.con} \affiliation{Department of mathematics,
North University of China,\\ Taiyuan 030051, P. R.China.}

\date{\today}

\begin{abstract}

A pulse vaccination SIR model with periodic infection rate $\beta
(t)$ have been proposed and studied. The basic reproductive number
$R_0$ is defined. The dynamical behaviors of the model are analyzed
with the help of persistence, bifurcation and global stability. It
has been shown that the infection-free periodic solution is globally
stable provided $R_0 < 1$ and  is unstable if $R_0>1$. Standard
bifurcation theory have been used to show the existence of the
positive periodic solution for the case of $R_0 \rightarrow1^+$.
Finally, the numerical simulations have been performed to show the
uniqueness and the global stability of the positive periodic
solution of the system.

\vspace{2mm} {\bf Key words:} \hspace{1mm} Epidemic model; Basic
reproductive number; Pulse vaccination; Uniform Persistence;
Globally stability; Infection-free periodic solution; Positive
periodic solution.
\end{abstract}

\maketitle

{\section*{ 1. Introduction}}

Transmissible diseases have tremendous influence in human life.
Every year billions of people suffer or die in various infectious
diseases. In recent times, the emerging and reemerging communicable
diseases have led to a revive interest in the study of infectious
disease. Mathematical models are widely used to understand the
mechanisms of spread of infectious diseases, and become an important
tool to analyze the spreading and controlling the diseases [1].
Epidemiology modelling and mathematical study on those model can
contribute to the design and analysis of epidemiological surveys,
suggest some crucial conclusion that should be collected. By
determining the key parameters, and finding it's effects of changes
in the parameter values, it identify the trends, make general
forecasts, and finally estimate the uncertainty in forecasts [2].
Another important motivation of the development of this field is the
evaluation of various vaccination/control strategies for human as
well as animal. Vaccination is an effective way to control the
transmission of a disease among the species. Mathematical modelling
can contribute to the design and assessment of the vaccination
strategies. A theoretical examination of pulse vaccination policy
for SIR model have been shown by  Stone et. al [3]. They found a
disease-free periodic solution with the same period as the pulse
vaccination, and studied the local stability of this solution. Nokes
and Swinton have studied the control of childhood viral infections
by pulse vaccination [4]. Major application of the pulse vaccination
method for SIR and SEIR epidemic models have been studied by
A.d'onofrio and his group [5-7,9].  They have studied global
asymptotic eradication in the presence of vaccine failure [5].
Yicang zhou and Hanwu Liu studied the stability of Periodic
Solutions for an SIS Model with Pulse Vaccination [8].  Fuhrman et.
al  studied asymptotic behavior of an SI Epidemic Model with Pulse
Removal [10].  The global stability of the disease-free periodic
solution for SIR and SIRS models with Pulse Vaccination have been
investigated by Jin [11].

Recently, pulse vaccination, the repeated application of vaccine
over a defined age range, is gained prominence as a strategy for the
elimination of childhood viral infectious such as measles and
poliomyelitis. This policy is based on the suggestion that measles
epidemics can be more efficiently controlled when the natural
temporal process of the epidemics is antagonized by another temporal
process [12,13].

On the other hand , in the last 30 years epidemiology of infectious
diseases has focused on the theoretical and the
numerical-experimental study of the effects of periodically varying
external factors on the time course of the incidence of infectious
diseases. It is well known that the epidemic models with constant
contact rate are not able to reproduce realistic endemic situations,
since incidence of many infectious diseases are not constant [14].
The endemic nature of many communicable diseases is characterized by
a wide range of temporal oscillatory patterns: annual or polyannual
periodicity [15], apparently random oscillations with a strong
annual component in the power spectrum [16] etc.

In this paper, we have studied  the dynamical behavior of a pulse
vaccination SIR model with periodic infection rate. The paper is
organized as follows: Section 2 gives the SIR model with pulse
vaccination and definition of the basic reproductive number $R_0$,
and demonstrates the existence of an infection-free periodic
solution. The global stability of the infection-free periodic
solution is obtained in Section 3.  In Section 4, we discuss the
persistence of the infectious disease by obtaining the uniform
persistence conditions $R_0>1$ of the infectious disease. Section 5,
concentrates on the existence of the positive periodic solution of
the periodic SIR model with pulse vaccination. The existence of
positive periodic solution is obtained  by using a well-known
bifurcation result of Rabinowitz [17] for $R_0 \rightarrow 1^+$.
Finally, numerical simulations and a brief discussion conclude the
paper.

\vspace{3mm} {\section*{ 2. The SIR model without  and  With pulse
 vaccination  }}

Our SIR model is based on the following assumptions:

 1. The total population size is constant and is denoted by $N$,
with $N=S+I+R$, the population is divided into three groups: (a) The
susceptible class, S, comprising those people who are capable of
catching the disease; (b) The infectives, I, comprising those who
are infected and capable of transmitting the disease; (c) The
recovered class, R, comprising those individuals who are immune.

\medskip
\par

 2. The total population size is N and the per capita birth rate
is a constant $b$. As births balance deaths we must have that the
per capita death rate is also $b$.

\medskip
\par

3. The population is uniform and mixes homogeneously.

\medskip
\par

 4. The infection rate $\beta(t)$ is defined as the total rate at
which potentially infectious contacts occur between two individuals.
A potentially infectious contact is one which will transmit the
disease if one individual is susceptible and the other is
infectious, so the total rate at which susceptibles become infected
is $\beta(t)SI$: Biological considerations mean that $\beta(t)$ is a
continuous function. We also assume that $\beta(t)$ is not
identically zero, positive, non-constant and periodic function with
period $\omega > 0$.

\medskip
\par

 5. The infectives move from the infectives class to the
recovered class at a constant rate $\gamma$ where $(1/\gamma$ ) is
the average infectious period conditional on survival to the end of
it.

\medskip
\par

 6. We also assume that $\alpha $ is a non-negative constant and
represents the death rate due to the disease (that is, the
disease-related death rate).

\medskip
\par

On the basis of the above assumptions a SIR model can be written as
a set of coupled non-linear ordinary differential equations as
follows:

$$
\left\{ \begin{array}{l}
S'=bK-\beta (t) SI- bS,\\
I'=\beta(t)SI -(b+\gamma+\alpha )I,\\
R'=\gamma I-b R.
\end{array}
\right. \eqno{(2.1)}
$$

Here the parameters $K, b, \gamma$ are all positive constants. The
constant $b K\equiv A$ is  the inflow rates, $b$ is the per capita
birth rate and the per capita death rate, so that $K$ represents a
carrying capacity, or maximum possible population size.

Adding the three equation of system (2.1), we get
$$
N'(t)=bK-b N-\alpha I. \eqno{(2.2)}
$$

For small value of the parameter $\alpha$ the total population size
become closer to the carrying capacity $K$.

According to usual convention, pulse vaccination can be defined as
the repeated application of vaccine across an age range. Let us
assume the pulse scheme proposes to vaccinate a fraction $p_k$ of
the entire susceptible population in a single pulse, applied $k$
time in time period $[0,\omega]$. Pulse vaccination gives lifelong
immunity to $p_k S$ susceptibles who are, as a consequence,
transferred to the recovered class ($R$) of the population.

When pulse vaccination is incorporated to SIR model (2.1), the model
takes the form as follows:
$$
\left\{
\begin{array}{ll}
S'=b K-\beta (t) SI-b S,& t\not= \tau_k.\\
I'=\beta (t) SI -(b+\gamma+\alpha )I,& k=0,1,2\cdots. \\
R'=\gamma I-b R,&
\end{array}
\right. \eqno{(2.3)}
$$

$$\left\{
\begin{array}{ll}
S(\tau^+_{k})=(1-p_k )S(\tau_k),& \\
I(\tau^+_{k})=I(\tau_{k}),& k=0,1,2\cdots.\\
R(\tau^+_{k})=R(\tau_{k})+p_k S(\tau_{k}).&
\end{array}
\right\}, \eqno{(2.4)}
$$

where $0\leq p_k<1 (k\in Z_+)$ are constants and $q>0$ is an integer
such that $p_{k+q}=p_k$,$\tau_{k+q}=\tau_k+\omega$. Note that the
dynamics of the total population size $N(t)$ still satisfy equation
(2.2).

From biological view point, we can assume that the domain
$$\Omega = \{(S, I,R) : S\geq 0, I\geq 0,R \geq 0,S+ I+R < K\}$$ is a positive
invariant set of system (2.3) and (2.4).

Let us start to analyze system (2.3) and (2.4) by  demonstrating the
existence of an infection-free periodic solution, in which
infectious individuals are entirely absent from the population
permanently i.e.,
$$
I(t)=0, t\geq 0. \eqno{(2.5)}
$$
Under this conditions, the systems (2.3) and (2.4) reduce to

$$
\left\{
\begin{array}{ll}
S'=b K-b S,& t\not= \tau_k.\\
R'=-b R,& k=0,1,2\cdots.\\
S(\tau^+_{k})=(1-p_k )S(\tau_k),& \\
R(\tau^+_{k})=R(\tau_{k})+p_k S(\tau_{k}).&
\end{array} \right.
\eqno{(2.6)}
$$
and $N'=bK-bN$, therefore $N(t)\rightarrow K$ as $t\rightarrow
\infty$. For convenient, we take $N=S+R=K$. Then, solving the
equation (2.6), we get
$$
S(t)=W(t,0)S(0)+bK\int_0^tW(t,\tau)d \tau$$
 and
$$
R(t)= W(t,0)R(0)+\Sigma_{0< \tau_k<t}W(t,\tau_k)p_k,
$$
where
$$ W(t,\tau)=\prod\limits_{\tau<
\tau_j<t}(1-p_j)e^{-b(t-\tau)}.
$$
Since $W(\omega,0)=\prod\limits_{j=1}^q(1-p_j)e^{-b\omega}<1$,
therefore equation (2.6) has a unique $\omega$-periodic solution
$(S^*(t),0,R^*(t))$ with the initial conditions
$S^*(0)=bK\int_0^{\omega}W(\omega,\tau)d\tau/ (1-W(\omega,0))$,
$R^*(0)=\sum\limits_{0<\tau_k<t}W(\omega,\tau_k)/ (1-W(\omega,0))$.

Let us define the basic reproductive rate of model (2.3) and (2.4)
as follows:

$$R_0=\frac{\int_0^\omega\beta(t)S^*(t)dt}{\omega(b+\alpha+\gamma)},
$$
where $S^*(t)$ is the periodic infection-free solution.

\vspace{3mm} {\section*{ 3.  Local and global asymptotic stability
of the infection free solution }}

In this section, we will prove the local and global asymptotic
stability of the infection free solution $(S^*(t),0,R^*(t))$.

The local stability of the $\omega$-period solution
$(S^*(t),0,R^*(t))$ may be determined by considering the linearized
SIR equation of (2.3), (2.4) about the known periodic solution
$(S^*(t),0,R^*(t))$ by setting $S(t) = S^*(t) + x(t)$, $I(t) =
y(t)$, $R(t) = R^*(t) + z(t)$, where $x(t), y(t)$, and $z(t)$ are
small perturbation. The linearized equations may be written as
$$
\left(
\begin{array}{l}
x(t)\\
y(t)\\
z(t)
\end{array} \right)
=\Phi (t)\left(
\begin{array}{l}
x(0)\\
y(0)\\
z(0)
\end{array} \right),
$$
where $\Phi (t) = \varphi_{ij}(t),i,j=1,2,3$ satisfies
$$
\frac{d\Phi (t)}{dt}=\left(
\begin{array}{ccc}
-b&-\beta(t)S^*(t)&0\\
0&\beta(t)S^*(t)-(b+\gamma+\alpha)&0\\
0&\gamma&-b
\end{array} \right)\Phi (t),
$$
with $\Phi(0) = E$, where $E $ is the identity matrix. Therefore,
the system ( 2.4) become

$$ \left(
\begin{array}{l}
x(\tau_k^+)\\
y(\tau_k^+)\\
z(\tau_k^+)
\end{array} \right)=
\left(\begin{array}{ccc}
1-p_k&0&0\\
0&1&0\\
p_k & 0 & 1
\end{array} \right)
\left(\begin{array}{l}
x(\tau_k)\\
y(\tau_k)\\
z(\tau_k)
\end{array} \right).
$$
Hence, according to the Floquet theory, if all eigenvalues of
$$
M=\left(\begin{array}{ccc}
\prod\limits_{k=1}^q(1-p_k)&0&0\\
0&1&0\\
p_q\prod\limits_{k=2}^q(1-p_k)&0&1
\end{array} \right)\Phi(\omega)
$$
are less than one, then the $\omega$-periodic solution
$(S^*(t),0,R^*(t))$ is locally stable. After calculation we get,

$$
\Phi (t)=\left (\begin{array}{ccc}
e^{-b t}&\varphi _{12}&0\\
0&\varphi _{22}&0\\
0 &\varphi _{32}&e^{-b t}
\end{array}
\right ),
$$
where

$$\begin{array}{lll}
\varphi _{22}&=&\exp(\int_{0}^{t}[\beta(\tau) S^*(\tau)-(b+\alpha +\gamma ]d\tau)\\

\vspace{2mm} \varphi _{12}&=&- \exp{(-b t)}\int_{0}^{t}\beta(\tau)S^*(\tau)\varphi _{22}(\tau )e^{b \tau}d\tau ,\\
\varphi _{32}&=&\exp{-b t}\int_{0}^{t}\gamma \varphi _{22}(\tau)e^{b
\tau}d\tau .
\end{array}$$

Thus, if $\mu_1 , \mu_2$ and $\mu_3$ are the eigenvalues of matrix
$M$, then they are  given by,
$$\mu_1=\prod\limits_{k=1}^{q}(1-p_k)e^{-b\omega},
\mu_3=e^{-b\omega}$$ and $$\mu_2=\varphi_{22}(\omega)=
\exp(\int_{0}^{t}\beta(\tau) S^*(\tau)d\tau-(b+\alpha +\gamma
)\omega.$$ Obviously $\mu_1<1$, $\mu_3<1$  and $\mu_2<1$ if and only
if
$$
\frac{\int_{0}^{\omega}\beta(\tau)
S^*(\tau)d\tau}{\omega}<b+\alpha+\gamma.
$$

Therefore, the periodic infection-free solution $(S^*(t), 0,R^*(t))$
is asymptotically stable if $R_0 < 1$. Now we are in a position to
summarize the above results in the following theorem:

{\bf Theorem 3.1.} The periodic infection-free solution $(S^*(t),
0,R^*(t))$ of system (2.3) and (2.4) is local asymptotically stable
provided $R_0 < 1$.

In order to prove  the global stability of infection-free solution
$(S^*(t), 0,R^*(t))$, we need the following lemma.

{\bf Lemma 3.1.} (Comparison theory [18]). Assume that $m\in
PC[R_+,R]$ with points of discontinuity at $t=\tau_n$ and is left
continuous at $t=\tau_n$, $n=1,2,\cdots$, and
$$\left\{
\begin{array}{l}
D_{-}m(t) \leq g(t,m(t)), t\neq\tau_n ,n=1,2,\cdots, \\
m(\tau_n^+)\leq \psi_n(m(\tau_n)), t=\tau_n ,n=1,2,\cdots,
\end{array}
\right.\eqno{(3.1)}
$$
where $g\in C[R_+\times R_+, R]$, $\psi_n\in C[R,R]$ and $\psi_n(u)$
is nondecreasing in $u$ for each $n=1,2,\cdots$. Let $r(t)$ be the
maximal solution of the scalar impulsive differential equation

$$\left\{
\begin{array}{l}
u' = g(t,m(t)), t\neq\tau_n ,n=1,2,\cdots, \\
u(\tau_n^+)= \psi_n(u(\tau_n)), t=\tau_n ,n=1,2,\cdots,\\
u(t_0^+)=u_0,
\end{array}
\right. \eqno{(3.2)}
$$
existing on $[t_0,\infty)$. Then $m(t_0^+)\leq u_0$ implies
$m(t)\leq r(t)$.

{\bf Remark.} In Lemma 3.1, assume the inequalities (3.1) reversed.
Let $\rho (t)$  be the minimal solution of $(3.2)$ existing on
$[t_0,\infty)$ . Then, $m(t)\geq \rho(t)$.

\vspace{3mm}
 {\bf Lemma 3.2.} For the following impulsive equation
$$
\left\{
\begin{array}{ll}
S'=b K-b S,& t\not= \tau_k.\\
S(\tau^+_{k})=(1-p_k )S(\tau_k),& n=1,2,\cdots,
\end{array}
\right. \eqno{(3.3)}
$$
has a unique positive $\omega$ -periodic solution $S^*(t, S^*_0)$
for which $S^*(0, S^*_0)= S^*_0$ and $S^*(t, S^*_0) > 0$, $t \in
R_+$, and $S^*(t, S^*_0)$ is global asymptotically stable in the
sense that $\lim\limits_{t\rightarrow \infty} |S(t,S_0)- S^*(t,
S^*_0)|=0$, where $S(t,S_0)$ is any solution of system (3.3) with
positive initial value $S(0,S_0)= S_0 > 0$.

 {\bf Proof.}  The first part of the proof follows from the results of Section  2.
 Therefore we have to show only
$\lim\limits_{t\rightarrow \infty} |S(t,S_0)- S^*(t, S^*_0)|=0$.

Since $$ |S(t,S_0)- S^*(t, S^*_0)|=W(t,0)|S_0-  S^*_0|,
$$
thus we need to show $W(t, 0)\rightarrow 0$ as $\rightarrow \infty$.
Suppose $t \in (n\omega , (n + 1)\omega]$, then
$$
\begin{array}{ll}
W(t,0)&=\prod\limits_{0< \tau_j<t}(1-p_j)e^{-bt}\\
&=\prod\limits_{0< \tau_j< n\omega}(1-p_j)e^{-bn\omega}
\prod\limits_{n\omega< \tau_j<t}(1-p_j)e^{-b(t-n\omega)} \\
&=[\prod\limits_{0<
\tau_j<\omega}(1-p_j)e^{-b\omega}]^n\prod\limits_{n\omega<
\tau_j<t}(1-p_j)e^{-b(t-n\omega)}\\
&\leq\prod\limits_{0< \tau_j<\omega}(1-p_j)[\prod\limits_{0<
\tau_j<\omega}(1-p_j)e^{-b\omega}]^n
\end{array}
$$
Thus $\lim\limits_{t\rightarrow \infty} W(t, 0) = 0$, since
$\prod\limits_{0< \tau_j<\omega}(1-p_j)e^{-b\omega}<1$. Hence the
proof.

Now, we are totally ready to show the global asymptotic stability
conditions of the infection free solution. We claim the following
theorem:

{\bf Theorem 3.2.}  The periodic infection-free solution $(S^*(t),
0,R^*(t))$  of system (2.3) and (2.4) is globally stable if $R_0 <
1$.

{\bf Proof.} From the first equation of (2.3) and (2.4), and by
using Lemma 3.1 and 3.2, we obtain that, for any given
$0<\varepsilon<\omega(1-R_0)(b+\gamma+\alpha)/ 2\int_0^{\omega}
\beta (t)dt$,  there exists $T_1>0$, such that
$$
S(t)<S^*(t)+\varepsilon, ~~~ \forall t>T_1. \eqno{(3.4)}
$$
By substituting (3.4) into second equation of (2.3), we obtain
$$
\frac{dI(t)}{dt}=\beta(t)SI-(b+\gamma+\alpha)I\leq
\beta(t)S^*(t)I(t)-(b+\gamma+\alpha)I(t)+\varepsilon\beta(t) I(t)
$$

By comparison theory, for $t\in (T_1+n\omega,T_1+(n+1)\omega]$, we
have
$$
\begin{array}{ll} I(t)&\leq
I(T_1)\exp(\int_{T_1}^{t}[\beta(\tau)S^*(\tau)-(b+\gamma+\alpha)+\beta(\tau)\varepsilon
])d\tau\\
&=I(T_1)\exp\{\int_{T_1}^{T_1+n\omega}[\beta(\tau)S^*(\tau)-(b+\gamma+\alpha)
]d\tau+\varepsilon\int_{T_1}^{T_1+n\omega}\beta(\tau)d\tau\\
&+\varepsilon\int_{T_1+n\omega}^{t}\beta(\tau)d\tau+\int_{T_1+n\omega}^{t}[\beta(\tau)S^*(\tau)-(b+\gamma+\alpha)
]d\tau\}\\
&\leq
I(T_1)\exp\{n\int_{0}^{\omega}[\beta(\tau)S^*(\tau)-(b+\gamma+\alpha)
]d\tau+(n+1)\varepsilon\int_{0}^{\omega}\beta(\tau)d\tau+\int_{0}^{\omega}\beta(\tau)S^*(\tau)d\tau\\
&\leq B
\exp\{n\omega(b+\gamma+\alpha)(R_0-1)+n\varepsilon\int_{0}^{\omega}\beta(\tau)d\tau\}\\
&\leq B\exp\{n\omega(b+\gamma+\alpha)\frac{1}{2}(R_0-1)\}
\end{array}
\eqno{(3.5)}
$$
where
$B=I(T_1)\exp\{\int_{0}^{\omega}\beta(\tau)(\varepsilon+S^*(t))d\tau\}$.
Thus if $R_0<1$, it follows from the above equation that
$\lim\limits_{t\rightarrow \infty}I(t)=0$.

 Let $ H(t) = |S(t) - S^*(t)|$, then we have

$$
D_+H(t)= \hbox{sign} (S(t) - S^*(t))(S'(t) - {S^*}'(t))\leq
-bH(t)+\beta(t)SI, \eqno{(3.6)}
$$
$$
H(\tau_k^+)=(1-p_k)H(\tau_k). \eqno{(3.7)}
$$
Since $S(t)\leq K$, $\beta(t)\leq \max\limits_{t\in
[0,\omega]}\beta(t)\equiv \beta^*$, the equation (3.6) reduces to

$$
D_+H(t)\leq -bH(t)+K\beta^*I(t). \eqno{(3.8)}
$$
For $t\in (T_1+n\omega,T_1+(n+1)\omega]$ , from (3.5), we see that
$I(t)\leq B e^{-\theta n}$, where
$\theta=\frac{1}{2}\omega(b+\gamma+\alpha)(1-R_0)$. By lemma 3.1 and
(3.7), (3.8), we have
$$
H(t)\leq H(T_1)\prod\limits_{T_1<
\tau_j<t}(1-p_j)e^{-b(t-T_1}+K\beta^*\int_{T_1}^t\prod\limits_{\tau<
\tau_j<t}(1-p_j)e^{-b(t-\tau)}I(\tau)d\tau. \eqno{(3.9)}
$$
Since
$$\begin{array}{l}
\int_{T_1}^t\prod\limits_{\tau<
\tau_j<t}(1-p_j)e^{-b(t-\tau)}I(\tau)d\tau\\
=\int_{T_1}^{\omega+T_1}\prod\limits_{\tau<
\tau_j<t}(1-p_j)e^{-b(t-\tau)}I(\tau)d\tau+\int_{\omega+T_1}^{2\omega+T_1}\prod\limits_{\tau<
\tau_j<t}(1-p_j)e^{-b(t-\tau)}I(\tau)d\tau+\cdots\\
 +\int_{(n-1)\omega+T_1}^{n\omega+T_1}\prod\limits_{\tau<
\tau_j<t}(1-p_j)e^{-b(t-\tau)}I(\tau)d\tau+\int_{n\omega+T_1}^{t}\prod\limits_{\tau<
\tau_j<t}(1-p_j)e^{-b(t-\tau)}I(\tau)d\tau\\
\leq B[\int_{T_1}^{\omega+T_1}\prod\limits_{\tau<
\tau_j<t}(1-p_j)e^{-b(t-\tau)}d\tau+\int_{\omega+T_1}^{2\omega+T_1}\prod\limits_{\tau<
\tau_j<t}(1-p_j)e^{-b(t-\tau)}e^{-\theta }d\tau+\cdots\\
 +\int_{(n-1)\omega+T_1}^{n\omega+T_1}\prod\limits_{\tau<
\tau_j<t}(1-p_j)e^{-b(t-\tau)}e^{-(n-1)\theta
}d\tau+\int_{n\omega+T_1}^{t}\prod\limits_{\tau<
\tau_j<t}(1-p_j)e^{-b(t-\tau)}e^{-\theta n}d\tau] \\
\leq
Be^{bT_1}(e^{b\omega}-1)e^{-bt}[\eta^{n-1}+\eta^{n-2}e^{b\omega-\theta}+\eta^{n-3}e^{2(b\omega-\theta)}+\cdots
+\eta e^{(n-1)(b\omega-\theta)}+e^{n(b\omega-\theta)}]/
b\\
=Be^{bT_1}(e^{b\omega}-1)e^{-bt}[\eta^n-e^{n(b\omega-\theta)}]/b(\eta-e^{(b\omega-\theta)})\\
\leq
B(e^{b\omega}-1)e^{-bn\omega}[\eta^n-e^{n(b\omega-\theta)}]/b(\eta-e^{(b\omega-\theta)}),
\end{array}
\eqno{(3.10)}
$$
where $0<\eta=\prod\limits_{j=1}^q(1-p_j)\leq1$. From (3.9) and
(3.10), we obtain that $\lim\limits_{t\rightarrow \infty}H(t)=0$,
i.e., $S(t) \rightarrow S^*(t)$ as $t\rightarrow \infty$. Similarly,
we can prove $R(t) \rightarrow R^*(t)$ as $t\rightarrow \infty$. The
proof is completed.

\vspace{3mm} {\section*{ 4.  The uniform persistence of the
infectious disease}}

In this section, we will discuss the uniform persistence of the
infectious disease, that is, $\lim\limits_{t\rightarrow \infty}\inf
I(t)>\alpha$ if $R_0>1$.

To discus the uniform persistence, we need a lemma first.

{\bf Lemma 4.1.} If $R_0>1$, then the disease uniformly weakly
persists in the population, in the sense that we will able to find a
constant $c>0$ such that $\lim\limits_{t\rightarrow \infty}\sup
I(t)>c$ for all solutions of (2.3) and (2.4).

{\bf Proof.}  Let us suppose that for a given $\varepsilon >0$,
there exists a solution with $\lim\limits_{t\rightarrow \infty}\sup
I(t)<\varepsilon$. From the first equation of (2.3), we have
$$
S'=b K-\beta (t) SI-b S \geq bK -\beta^* K \varepsilon-bS, t\not=
\tau_k.
$$

Consider the following equation
$$\left\{\begin{array}{l}
u'= bK -\beta^* K \varepsilon-bu, t\not= \tau_k.\\
u(\tau^+_k)=(1-p_k)u(\tau_k), k=1,2,\cdots
\end{array}
\right.\eqno{(4.1)}
$$

By lemma 3.2., the (4.1) has a unique  positive $\omega$ -periodic
solution $u^*(t)$ for which $u^*(0)= u^*_0$ and
 $u^*(t)$ is global asymptotically stable. So $$
 S^*(t)-u^*(t)=\beta^*K\varepsilon[\frac{W(t,0)\int_0^{\omega}W(\omega,\tau)d\tau}{1-W(\omega,0)}+\int_0^{t}
 W(t,\tau)d\tau].
 \eqno{(4.2)}
 $$
Let $$ \Delta=\beta^*K \max\limits_{0\leq t\leq
\omega}\{\frac{W(t,0)\int_0^{\omega}W(\omega,\tau)d\tau}{1-W(\omega,0)}+\int_0^{t}
 W(t,\tau)d\tau\}.
 $$
By (4.2), we see that
$$u^*(t)\geq S^*(t)-\Delta \varepsilon.
\eqno{(4.3)}$$.

By comparison theory, we obtain that
$$\begin{array}{ll}
I'&=\beta (t) SI -(b+\gamma+\alpha )I\\
&\geq I(t)[\beta (t)u(t)-(b+\gamma+\alpha)].
\end{array}
\eqno{(4.4)}
$$
Since, $u^*(t)$ is global asymptotically stable, for our previous
$\varepsilon$ above, there exist $T_1>0$, such that $u(t)\geq
u^*(t)-\varepsilon$, $t>T_1$. From (4.3) and (4.4), we get
$$
I'(t)\geq I(t)[\beta (t)S^*(t)-(b+\gamma+\alpha
)-\varepsilon(1+\Delta)\beta (t]. \eqno{(4.5)}
$$
Integrating over intervals $[T_1, t]$ we obtain
$$
\begin{array}{ll} I(t)&\geq
I(T_1)\exp(\int_{T_1}^{t}[\beta(\tau)S^*(\tau)-(b+\gamma+\alpha)-\varepsilon(1+\Delta)\beta(t)
]d\tau\\
&=I(T_1)\exp\{\int_{T_1}^{T_1+n\omega}[\beta(\tau)S^*(\tau)-(b+\gamma+\alpha)
]d\tau-\varepsilon
(1+\Delta)\beta(t)n\omega\\
&+\int_{T_1+n\omega}^{t}[\beta(\tau)S^*(\tau)-(b+\gamma+\alpha)
]d\tau-\varepsilon(1+\Delta)\beta(t)(t-n\omega)\}\\
&\geq C
\exp\{n\omega(b+\gamma+\alpha)(R_0-1)-\varepsilon(1+\Delta)\beta^*n\omega\}.
\end{array}
\eqno{(4.6)}
$$
where $t\in (T_1+n\omega, T_1+(n+1)\omega]$,
$C=I(T_1)\exp[-(b+\gamma+\alpha+\varepsilon(1+\Delta)\beta^*)\omega]$.
Taking $$ 0<\varepsilon\leq
\frac{(b+\gamma+\alpha)(R_0-1)}{2\beta^*(1+\Delta)},
$$
thus $I(t)\rightarrow \infty$ as $t\rightarrow \infty$, a
contradiction to the fact that $I(t)$ is bounded. This finishes the
proof.

{\bf Theorem 4.1.} If $R_0>1$, then the disease uniformly
persistence, it is that there exists a positive constant $\sigma$
such that for every positive solution of (2.3) and (2.4),
$$\lim\limits_{t\rightarrow \infty}\inf I(t)\geq\sigma>0.$$

{\bf Proof.}  Let
$$
0<\eta \leq \frac {bK(R_0-1)}{2R_0(\alpha+\gamma)}.
$$
It can be obtained  from the Lemma 4.1  that for any positive
solution of (2.3) and (2.4) there exists at least one $t_0 > 0$ such
that $I(t_0)>\eta>0$. Then, we are left to consider two
possibilities. The first case is $I(t)\geq\eta$ for all large $t\geq
t_0$. The second one is $I(t)$ oscillates about $\eta$ for large t.
The conclusion of Theorem 4.1 is obvious in the first case since we
can choose $\sigma=\eta$.

For the second case, let $t_1 > t_0$ and $t_2 > t_1$ satisfy
$$
I(t_1)=I(t_2)=\eta, \ \ \ \hbox{and} \ \ \ I(t)<\eta \ \ \
\hbox{for}\ \ \ t_1<t<t_2.
$$

Next, we introduce a new variable $V = S + I$, and it follows from
the first two equations of (2.3) and (2.4) that

$$
\left\{
\begin{array}{ll}
V'=b K-bV-(\gamma+\alpha)I,& t\not= \tau_k.\\
I'=\beta (t) (V-I)I -(b+\gamma+\alpha )I,& k=0,1,2\cdots. \\
\end{array}
\right. \eqno{(4.7)}
$$

$$
\left\{
\begin{array}{ll}
V(\tau^+_{k})=(1-p_k )V(\tau_k)+p_kI(\tau_k),& \\
I(\tau^+_{k})=I(\tau_{k}),& k=0,1,2\cdots.\\
\end{array}
\right. \eqno{(4.8)}
$$

If $I(t)\leq \eta$, then
$$
\left\{
\begin{array}{ll}
V'=b K-bV-(\gamma+\alpha)I>\frac{bK}{2}(1+\frac{1}{R_0})-bV,& t\not= \tau_k.\\
V(\tau^+_{k})>(1-p_k )V(\tau_k)).
\end{array}
\right. \eqno{(4.9)}
$$

Consider the following equation
$$
\left\{
\begin{array}{ll}
x'=\frac{bK}{2}(1+\frac{1}{R_0})-bx,& t\not= \tau_k.\\
x(\tau^+_{k})=(1-p_k )x(\tau_k).
\end{array}
\right. \eqno{(4.10)}
$$
By using Lemma 3.2, we can obtain equation (4.10) has a unique
positive $\omega$- periodic solution $x^*(t)$, and $x^*(t)$ is
global asymptotically stable in the sense that
$\lim\limits_{t\rightarrow \infty} |x(t,x_0)- x^*(t)|=0$, where
$x(t,x_0)$ is any solution of system (4.10) with positive initial
value $x(0,x_0)= x_0 > 0$. By (3.3) and (4.13) we can easily get
$$
x^*(t)=\frac{1}{2}(1+\frac{1}{R_0})S^*(t)
$$
where $S^*(t)$ is a unique positive $\omega$- periodic solution of
system (3.3).

Then the comparison principle and the global asymptotically stable
of $x^*(t)$ implies that there exists a positive constant $T_1>0$,
such that
$$\begin{array}{l}
V(t)>\frac{1}{2}(1+\frac{1}{R_0})S^*(t),\ \ \ \hbox{for all}\ \ \
t>t_1+T_1.
\end{array}
\eqno{(4.11)}
$$

For above $t_1$ and $\eta$ the solution of the initial value problem
$$
y'(t)=\beta(t)[\frac{1}{2}(1+\frac{1}{R_0})S^*(t)-y]y-(b+\gamma+\alpha
)y,\ \ \ \ y(t_1)=\eta \eqno{(4.12)}
$$
is
$$
y(t)=\frac{\eta\exp(\int_{t_1}^t[\frac{1}{2}(1+\frac{1}{R_0})\beta(\tau)S^*(\tau)-(b+\gamma+\alpha
)]d\tau)}{1+\eta\int_{t_1}^t\beta(\tau)\exp(\int_{t_1}^\tau[\frac{1}{2}(1+\frac{1}{R_0})\beta(\theta)S^*(\theta)-(b+\gamma+\alpha
)]d\theta)d\tau} \eqno{(4.13)}$$

The differential equation in (4.12) has a periodic solution
$$
y^*(t)=\frac{y^*\exp(\int_{t_1}^t[\frac{1}{2}(1+\frac{1}{R_0})\beta(\tau)S^*(\tau)-(b+\gamma+\alpha
)]d\tau)}{1+y^*\int_{t_1}^t\beta(\tau)\exp(\int_{t_1}^\tau[\frac{1}{2}(1+\frac{1}{R_0})\beta(\theta)S^*(\theta)-(b+\gamma+\alpha
)]d\theta)d\tau} \eqno{(4.14)}$$

where $
y^*=\frac{\exp(\int_{t_1}^{t_1+\omega}[\frac{1}{2}(1+\frac{1}{R_0})\beta(\tau)S^*(\tau)-(b+\gamma+\alpha
)]d\tau)-1}{\int_{t_1}^{t_1+\omega}\beta(\tau)\exp(\int_{t_1}^\tau[\frac{1}{2}(1+\frac{1}{R_0})\beta(\theta)S^*(\theta)-(b+\gamma+\alpha
)]d\theta)d\tau} $. Since the  condition $R_0 > 1$ of Theorem 4.1
implies that
$\int_{0}^{\omega}[\frac{1}{2}(1+\frac{1}{R_0})\beta(\tau)S^*(\tau)-(b+\gamma+\alpha
)]d\tau=\frac{\omega(b+\gamma+\alpha)(R_0-1)}{2}>0$,  from the
limits
$$
\lim\limits_{t\rightarrow
\infty}\exp(\int_{t_1}^t[\frac{1}{2}(1+\frac{1}{R_0})\beta(\tau)S^*(\tau)-(b+\gamma+\alpha
)]d\tau)=\infty,
$$
$$
\lim\limits_{t\rightarrow
\infty}\int_{t_1}^t\beta(\tau)\exp(\int_{t_1}^\tau[\frac{1}{2}(1+\frac{1}{R_0})\beta(\theta)S^*(\theta)-(b+\gamma+\alpha
)]d\theta)d\tau=\infty
$$
and the expression of $y(t)-y^*(t)$ it follows that
$$\lim\limits_{t\rightarrow
\infty}|y(t)-y^*(t)|=0. \eqno{(4.15)}
$$
From (4.15) we see that there exists a positive constant $T_2 > 0$
such that
$$\begin{array}{l}
y(t)>\rho \equiv\frac{1}{2}\min\limits_{t_1\leq t\leq
t_1+\omega}y^*(t)>0,\ \ \ \hbox{for all}\ \ \ t>t_1+T_2.
\end{array}
\eqno{(4.16)}
$$

Let $T=\max\{T_1,T_2\}$ and define
$$
\sigma=\min\{\rho, \eta\exp(-(b+\gamma+\alpha)T\}.
$$

If $t_2-t_1 < T$, from the following  equation

$$
\begin{array}{ll}
I'=\beta (t) (V-I)I -(b+\gamma+\alpha )I,&
\end{array}
\eqno{(4.17)}
$$
we have the inequality
$$ I'(t)> -(b+\gamma+\alpha )I,
$$
and the comparison principle implies that $ I(t)\geq
\eta\exp\{-(b+\gamma+\alpha )\}\geq
\eta\exp\{-(b+\gamma+\alpha)T\}$, i.e., $I(t)\geq \sigma$ for all
$t\in (t_1, t_2)$. If $t_2-t_1> T$, we divide the interval $[t_1,
t_2]$ into two subintervals $[t_1, t_1 +T]$ and $[t_1 +T, t_2]$.
$I(t) \geq \sigma$ is obvious in the interval $[t_1, t_1 + T]$. By
(4.11) and (4.17) we see that in the interval $[t_1 + T, t_2]$,
$$
I'(t)\geq\beta(t)[\frac{1}{2}(1+\frac{1}{R_0})S^*(t)-I]I-(b+\gamma+\alpha
)I.  \eqno{(4.18)}
$$
Therefore again by using comparison principle  we get $I(t) \geq
y(t) \geq \rho\geq\sigma$ for $t\in [t_1 + T, t_2]$. The above
analysis is independent of any interval $[t_1, t_2]$, and the choice
of $\sigma$ is independent of any positive solution of (2.3) and
(2.4). Therefore, the persistence is uniform for all positive
solution.

\vspace{5mm} {\section*{ 5.  Existence of the positive
$\omega$-periodic solution and bifurcation}}

Let $PC(J,R)$ ($J\subset R$) be the set of continuous functions
$\psi: J\rightarrow R$ for $t\in J, t\neq \tau_k$, that have
discontinuities of the first kind at the points $\tau_k\in J$ where
they are continuous from the left. Let $PC'(J,R)$ be the set of
functions $\psi: J\rightarrow R$ with derivative
$\frac{d\psi}{dt}\in PC(J,R)$. We consider the Banach spaces of
$\omega$-periodic functions $PC_{\omega}=\{\psi\in PC([0,\omega],R)|
\ \psi(0)=\psi(\omega)\}$ under the supremum norm
$\parallel\psi\parallel_{PC_{\omega}}=\sup\{\mid\psi\mid : t\in
[0,\omega]\}$, and $PC'_{\omega}=\{\psi\in PC'([0,\omega],R)| \
\psi(0)=\psi(\omega)\}$ under the supremum norm
$\parallel\psi\parallel_{PC'_{\omega}}=\max\{\parallel\psi\parallel_{PC_{\omega}},
\parallel \psi' \parallel_{PC'_{\omega}}
\}$. We will also consider the product space $PC_{\omega}\times
PC_{\omega}$ which is also a Banach space under the norm
$\parallel(\psi_1,\psi_1)\parallel_{PC_{\omega}}=\parallel\psi_1\parallel_{PC_{\omega}}+\parallel\psi_2\parallel_{PC_{\omega}}$.
Moreover, for any $f\in C_{\omega}$ (or $PC_{\omega}$) we define the
average of $f$ by $\bar{f}: =\int_0^{\omega}f(s)ds / \omega$.

We need the following lemmas which are about linear impulsive
periodic equations (also see [20,21] ).

{\bf Lemma 5.1.} Suppose $c_{ij}(t)\in PC_{\omega}$, and

(1). if $\bar{c}_{11}\neq \frac{1}{\omega} \ln
[\prod\limits_{k=1}^{q}\frac{1}{1-p_k}]$ and $\bar{c}_{22}\neq 0$,
then the linear homogeneous periodic impulsive equation
$$
\left\{
\begin{array}{ll}
x_1'=c_{11}x_1(t)+c_{12}x_2(t),&  t\not= \tau_k\\
x_2'=c_{22}x_2(t),& k=0,1,2\cdots. \\
x_1(\tau^+_{k})=(1-p_k )x_1(\tau_k),& t=\tau_k\\
x_2(\tau^+_{k})=x_2(\tau_{k}),
\end{array}
\right. \eqno{(5.1)}
$$
has no nontrivial solution in $PC_{\omega}\times PC_{\omega}$. In
this case, the nonhomogeneous system
$$
\left\{
\begin{array}{ll}
y_1'=c_{11}y_1(t)+c_{12}y_2(t)+f_1,&  t\not= \tau_k\\
y_2'=c_{22}y_2(t)+f_2,& k=0,1,2\cdots. \\
y_1(\tau^+_{k})=(1-p_k )y_1(\tau_k),& t=\tau_k\\
y_2(\tau^+_{k})=y_2(\tau_{k}),
\end{array}
\right. \eqno{(5.2)}
$$
has a unique solution $(y_1,y_2)\in PC_{\omega}\times PC_{\omega}$
for every $(f_1,f_2)\in PC_{\omega}\times PC_{\omega}$ and the
operator $L: PC_{\omega}\times PC_{\omega}\rightarrow
 PC_{\omega}\times PC_{\omega}$ defined by $(y_1,y_2)= L (f_1,
 f_2)$ is linear and relatively compact.

 (2). if $\bar{c}_{11}\neq \frac{1}{\omega} \ln
[\prod\limits_{k=1}^{q}\frac{1}{1-p_k}]$ and $\bar{c}_{22}=0$, then
(5.1) has exactly one independent solution in $PC_{\omega}\times
PC_{\omega}$.

{\bf Proof.}  (1) Since
$$
x_2(t)=x_2(0)\exp \int_0^{t}c_{22}(s)ds \eqno{(5.3)}
$$
the condition $\bar{c}_{22}\neq 0$ implies that $x_2\notin
PC_{\omega}$ unless $x_2\equiv 0$. Then
$x_1(t)=\prod\limits_{0<\tau_k<t}(1-p_k)x_1(0)\exp\int_0^tc_{11}(s)ds
$ and $\bar{c}_{11}\neq \frac{1}{\omega} \ln
[\prod\limits_{k=1}^{q}\frac{1}{1-p_k}]$ in turn implies $x_1\notin
PC_{\omega}$ unless $x_1\equiv 0$.

In this case,

$$
y_2'=c_{22}y_2(t)+f_2
$$
has a unique solution $y_2(t)\in PC_{\omega}$ and the operator $L_2:
PC_{\omega}\rightarrow PC_{\omega}$ defined by $y_2\equiv L_2 f$ is
linear and relatively compact. Furthermore,

$$
\left\{
\begin{array}{ll}
y_1'=c_{11}y_1(t)+f_3,&  t\not= \tau_k,\\
y_1(\tau^+_{k})=(1-p_k )y_1(\tau_k),& t=\tau_k,
\end{array}
\right.
$$
for $f_3=c_{12}L_2f_2+f_1\in PC_{\omega}$ has a unique solution
(since $\bar{c}_{11}\neq \frac{1}{\omega} \ln
[\prod\limits_{k=1}^{q}\frac{1}{1-p_k}])$ in $PC_{\omega}$ and
$y_1=L_1 f_3$ defines a linear, relatively compact operator $$L_1:
PC_{\omega}\rightarrow PC_{\omega}.$$

Thus (5.2) has a unique $\omega$-period solution in
$PC_{\omega}\times PC_{\omega}$ given by $(y_1,y_2)=L(f_1,f_2)$,
where
$$
L(f_1,f_2)=(L_1(c_{12}L_2f_2)+f_1),L_2f_2).\eqno{(5.4)}
$$

(2). Under the stated assumptions, $x_2$ as given in (5.3) lies in
$PC_{\omega}$ for all initial conditions $x_2(0)$. Now if
$\bar{c}_{11}\neq \frac{1}{\omega} \ln
[\prod\limits_{k=1}^{q}\frac{1}{1-p_k}]$, then
$$
\left\{
\begin{array}{ll}
x_1'=c_{11}y_1(t)+c_{12}x_2(0)exp\int_0^tc_{22}(s)ds,&  t\not= \tau_k,\\
x_1(\tau^+_{k})=(1-p_k )x_1(\tau_k),& t=\tau_k,
\end{array}
\right.
$$
has a unique solution in $PC_{\omega}$.

{\bf Lemma 5.2.[18]} Suppose $a\in PC_{\omega}$ and $\bar{a}=
\frac{1}{\omega} \ln [\prod\limits_{k=1}^{q}\frac{1}{1+c_k}]$, then
$$\left\{
\begin{array}{ll}
z'=az+f,&  t\not= \tau_k,\\
z(\tau^+_{k})=(1+c_k )z(\tau_k),& t=\tau_k,
\end{array}
\right. \eqno{(5.5)}
$$

has a solution $z\in PC_{\omega}$ if and only if
$$
\int_0^\omega\prod\limits_{0<\tau_k<t}\frac{1}{1+c_k}\exp[-\int_0^ta(u)du]fdt=0
$$

For the existence of positive periodic solution, we have the
following theorem:

{\bf Theorem 5.1.} Let us assume $\beta(t)\in PC_{\omega}$, then
 there exists a sufficiently small constant $c_0>0$, such
 that for each $\beta(t)\in PC_{\omega}$ satisfying $b+\alpha+\gamma<\overline{\beta(t)
 S^*(t)}<b+\alpha+\gamma+c_0$, there exists a solution $(S(t),
 I(t))\in PC_{\omega}\times PC_{\omega}$ of (2.3) and (2.4) satisfying $S(t)<S^*(t), I(t)>0$ for all $t$.

{\bf Proof.} Let $u_1=S(t)-S^*(t), u_2=I(t)$ in (2.3) and (2.4),
then we have

$$
\left\{
\begin{array}{ll}
u_1'=-bu_1-\beta(t)S^*(t)u_2-\beta(t)u_1u_2,&  t\not= \tau_k\\
u_2'=[\beta(t)S^*(t)-(b+\gamma+\alpha)]u_2+\beta(t)u_1u_2 ,& k=0,1,2\cdots. \\
u_1(\tau^+_{k})=(1-p_k )u_1(\tau_k),& t=\tau_k\\
u_2(\tau^+_{k})=u_2(\tau_{k}),
\end{array}
\right. \eqno{(5.6)}
$$
Denote $\theta(t)=\beta(t)S^*(t)-\mu$, $\mu= \overline{\beta(t)
S^*(t)}$, then $\beta(t)S^*(t)=\theta(t)+\mu$ and
$\overline{\theta(t)}=0$, thus (5.6) becomes
$$
\left\{
\begin{array}{ll}
u_1'=-bu_1-\beta(t)S^*(t)u_2-\beta(t)u_1u_2,&  t\not= \tau_k\\
u_2'=[\theta(t) -(b+\gamma+\alpha)]u_2+\mu u_2+\beta(t)u_1u_2 ,& k=0,1,2\cdots. \\
u_1(\tau^+_{k})=(1-p_k )u_1(\tau_k),& t=\tau_k\\
u_2(\tau^+_{k})=u_2(\tau_{k}).
\end{array}
\right. \eqno{(5.7)}
$$
Consider the linear homogeneous system
$$
\left\{
\begin{array}{ll}
v_1'=-bv_1-\beta(t)S^*(t)v_2,&  t\not= \tau_k\\
v_2'=[\theta(t) -(b+\gamma+\alpha)]v_2,& k=0,1,2\cdots. \\
v_1(\tau^+_{k})=(1-p_k )v_1(\tau_k),& t=\tau_k\\
v_2(\tau^+_{k})=v_2(\tau_{k}).
\end{array}
\right. \eqno{(5.8)}
$$
Notice that
$$
\begin{array}{l}
\overline{c_{22}}=\overline{\theta(t)
-(b+\gamma+\alpha)}=-(b+\gamma+\alpha)\neq 0,\\
\overline{c_{11}}=-b\neq \frac{1}{\omega} \ln
[\prod\limits_{k=1}^{q}\frac{1}{1-p_k}]>0,
\end{array}
$$
and so (5.8) satisfies the hypotheses of part (1) of Lemma 5.1.
Consequently we have the compact linear operator $L:
PC_{\omega}\times PC_{\omega}\rightarrow
 PC_{\omega}\times PC_{\omega}$ given by (5.4). Using $L$ we can
 equivalently write system (5.7) as the operator equation
 $$
 (u_1,u_2)=\mu L^*(u_1,u_2)+G(u_1,u_2),
 \eqno{(5.9)}
 $$
 where
 $$
\begin{array}{l}
L^*(u_1,u_2)=(L_1(-\beta(t)S^*(t)L_2u_2), L_2u_2),\\
G(u_1,u_2)=(L_1(-\beta(t)S^*(t)L_2g_2(u_1,u_2)+g_1(u_1,u_2)),
L_2g_2),
\end{array}
$$
and $g_1(u_1,u_2)=-\beta(t)u_1u_2$, $g_2(u_1,u_2)=\beta(t)u_1u_2$.
Here $ L: PC_{\omega}\times PC_{\omega}\rightarrow
 PC_{\omega}\times PC_{\omega}$ is linear and relatively compact, and
 $G:
PC_{\omega}\times PC_{\omega}\rightarrow
 PC_{\omega}\times PC_{\omega} $
is quasiequicontinuous and relatively compact  and satisfies $G =
o(\parallel (u_1, u_2)\parallel_{PC_{\omega}})$ near $(0, 0)$. Note
that we have already got the operator equation (5.9) in such a form
so that now we can apply standard bifurcation theorems and
techniques directly to this equation. A non-trivial solution $(u_1,
u_2)\neq (0, 0)$ of (5.9) in $ PC_{\omega}\times PC_{\omega}$, for
some $\mu \in R $ ($R$ is the set of the reals), yields a solution
$(S,I) = (u_1 +S^*, u_2) $ of the system (2.3) and (2.4) for $ \mu
=\overline{\beta(t) S^*(t)} $. Solution $(S,I)\neq (S^*,0) $ will be
called non-trivial solutions of (2.3) and (2.4).

To prove Theorem 5.1, we are going to apply some well-known but
pretty standard local bifurcation techniques to the operator
equations (5.9)[17]. The bifurcation can occur only at the
non-trivial solutions of the linearized problem
$$
 (v_1,v_2)=\mu L^*(v_1,v_2), \ \ (v_1,v_2)\neq (0,0), \ \ \mu \in
 R.
 \eqno{(5.10)}
 $$

Let $(v_1,v_2)\in PC_{\omega}\times PC_{\omega}$ be a solution of
equation (5.10) for some $\mu \in R$, then by the very manner in
which $L^*$ is defined, $(v_1,v_2)$ satisfies the system

$$
\left\{
\begin{array}{ll}
v_1'=-bv_1-\beta(t)S^*(t)v_2,&  t\not= \tau_k\\
v_2'=[\theta(t)+\mu -(b+\gamma+\alpha)]v_2,& k=0,1,2\cdots. \\
v_1(\tau^+_{k})=(1-p_k )v_1(\tau_k),& t=\tau_k\\
v_2(\tau^+_{k})=v_2(\tau_{k}).
\end{array}
\right. \eqno{(5.11)}
$$
and conversely. By using Lemma 5.1, we see that (5.11) and hence
(5.10) has a non-trivial solution in $PC_{\omega}\times PC_{\omega}$
if and only if $\mu=\mu^*$ where
$$
\mu^*=b+\gamma+\alpha,
$$
If $\mu=\mu^*$, then, by part (2) of Lemma 5.1, (5.11) has one
independent solution in $ PC_{\omega}\times PC_{\omega}$. Therefore
a well-known result [17, 20-21] says that bifurcation occurs at this
simple eigenvalue; hence there exists a continuum $ U =\{(\mu,
u_1,u_2)\}\subseteq R \times PC_{\omega}\times PC_{\omega} $ of
non-trivial solutions of (5.9) such that the closure $ \overline{U}$
of $U$ contains $(\mu^*, 0, 0)$. This continuum gives rise to a
continuum $C = \{(\mu;S,I)\} \subseteq R \times PC_{\omega}\times
PC_{\omega} $ of non-trivial solutions of (2.3) and (2.4) whose
closure $\overline{C}$ contains the bifurcation point $(\mu^*; S^*,
0)$.

To observe that particular solutions in C correspond to solutions
$(S,I)$ of (2.3) and (2.4) with the properties described in Theorem
3.1, we investigate the nature of the continuum $C$ near the
bifurcation point $(\mu^*, 0, 0)$ by expanding $\mu$ and $(u_1,
u_2)$ in the Lyapunov-Schmidt series (for small $\varepsilon$) [22]:
$$
\mu=\mu^*+\mu_1\varepsilon+\cdots, \ \
u_i=u_{i1}\varepsilon+u_{i2}\varepsilon^2+\cdots, i=1,2
$$
for $u_{ij}\in PC_{\omega}$. We substitute these series into the
system (5.7) and equate the coefficients of $\varepsilon$ and
$\varepsilon^2$, we find
$$
\left\{
\begin{array}{ll}
u'_{11}=-bu_{11}-\beta(t)S^*(t)u_{21},&  t\not= \tau_k\\
u'_{21}=[\theta(t)+\mu^* -(b+\gamma+\alpha)]u_{21},& k=0,1,2\cdots. \\
u_{11}(\tau^+_{k})=(1-p_k )u_{11}(\tau_k),& t=\tau_k\\
u_{21}(\tau^+_{k})=u_{21}(\tau_{k}).
\end{array}
\right. \eqno{(5.12)}
$$
and
$$
\left\{
\begin{array}{ll}
u'_{12}=-bu_{12}-\beta(t)S^*(t)u_{22}-\beta(t)u_{11}u_{21},&  t\not= \tau_k\\
u'_{22}=[\theta(t)+\mu^* -(b+\gamma+\alpha)]u_{22}+\mu_1 u_{21}+\beta(t)u_{11}u_{21} ,&  \\
u_{12}(\tau^+_{k})=(1-p_k )u_{12}(\tau_k),& t=\tau_k\\
u_{22}(\tau^+_{k})=u_{22}(\tau_{k}), \ k=0,1,2\cdots.&
\end{array}
\right. \eqno{(5.13)}
$$
Obviously $(u_{11}, u_{21})\in PC_{\omega}\times PC_{\omega} $ must
be a solution of (5.11). We choose the specific solution satisfying
the initial conditions $u_{21}(0)=1$. Then
$$ u_{21}(t)=\exp(\int_0^t[\theta(s)+\mu^*
-(b+\gamma+\alpha)]ds)>0. \eqno{(5.14)}
$$
Moreover, $u_{11}(t)$ is the $\omega$-periodic solution of the
linear equation
$$
\left\{
\begin{array}{ll}
u'_{11}=-bu_{11}-\beta(t)S^*(t)u_{21},&  t\not= \tau_k\\
u_{11}(\tau^+_{k})=(1-p_k )u_{11}(\tau_k),& t=\tau_k
\end{array}
\right. \eqno{(5.15)}
$$
Hence $$ u_{11}=-\int_0^{\omega}G(t,s)\beta(s)S^*(s)u_{21}(s)ds,
$$
Where $G(t,s)$ is Green's function given by
$$
G(t,s)=\left\{
\begin{array}{ll}
X(t)(1-X(\omega))^{-1}/ X(s),&  0\leq s \leq t\leq\omega,\\
X(t+\omega)(1-X(\omega))^{-1}/ X(s),&  0\leq t \leq s\leq\omega,\\
G(t-k\omega,s-j\omega),& j\omega<s<j\omega,k\omega<t<k\omega+\omega,
\end{array}
\right.
$$
and
$$
X(t)=\prod\limits_{0<\tau_k<t}(1-p_k)e^{-bt}.
$$
Note that $u_{12}(t)>0$ for all $t$, and also since
$$
X(\omega)=\prod\limits_{k=1}^q(1-p_k)e^{-b\omega}<1,
$$
therefore Green's function $G(t,s)>0$ and thus $u_{11}(t)<0$ for all
$t$. Using Lemma 5.2 on the second equation of (5.13), we obtain
$$
\int_0^{\omega}u_{21}(t)(\mu_1
+\beta(t)u_{11})\exp(\int_0^{t}[\theta(s)+\mu^*
-(b+\gamma+\alpha)]ds)=0.\eqno{(5.16)}
$$
From the equation of (5.16), we obtain
$$ \mu_1
=-\overline{\beta(t)u_{11}(t)}>0.
$$

Thus we conclude (see [17]) that near the bifurcation point $(\mu^*;
0, 0)$ (say, for a sufficiently small constant $c_0
>0$, $\mu^*<\mu  < \mu^*+c_0$ ), there is a  branch $
U^+=\{(\mu; u_1,u_2)\in U: \mu^*\leq\mu<\mu^*+c_0,
u_1(t)<0,u_2(t)>0\}$.

Therefore, we have the piecewise continuous branch $C^+$ of
solutions of the from $(\mu;S,I)\in R\times PC_{\omega}\times
PC_{\omega}$, satisfying $S(t)<S^*(t), I(t)>0$ for all $t$. Since
$\mu^*\leq\mu<\mu^*+c_0 $ is equivalent to
$b+\alpha+\gamma<\overline{\beta(t) S^*(t)}<b+\alpha+\gamma+c_0$,
which again equivalent to $R_0\rightarrow 1^+$. Hence the proof.

\vspace{3mm} {\section *{6. \ Numerical  Simulation   }}

In our proposed pulse vaccination SIR model with periodic infection
rate $\beta (t)$  we obtain that the periodic infection-free
solution is globally stable provided the basic reproductive number
$R_0$ is less than 1. On the other hand, periodic infection-free
solution is unstable if the basic reproductive number is greater
than 1 and in this case the disease will persist in the population.
Moreover, the positive periodic solution exists when $R_0\rightarrow
1^+$.

In this section we have performed numerical simulation to show the
geometric impression of our results. In all simulations, the units
are set to unity (scaled to unity).

\begin{minipage}[c]{1\textwidth}
  \hspace{-1cm}
 \subfigure[]{\includegraphics[width=0.33\textwidth]{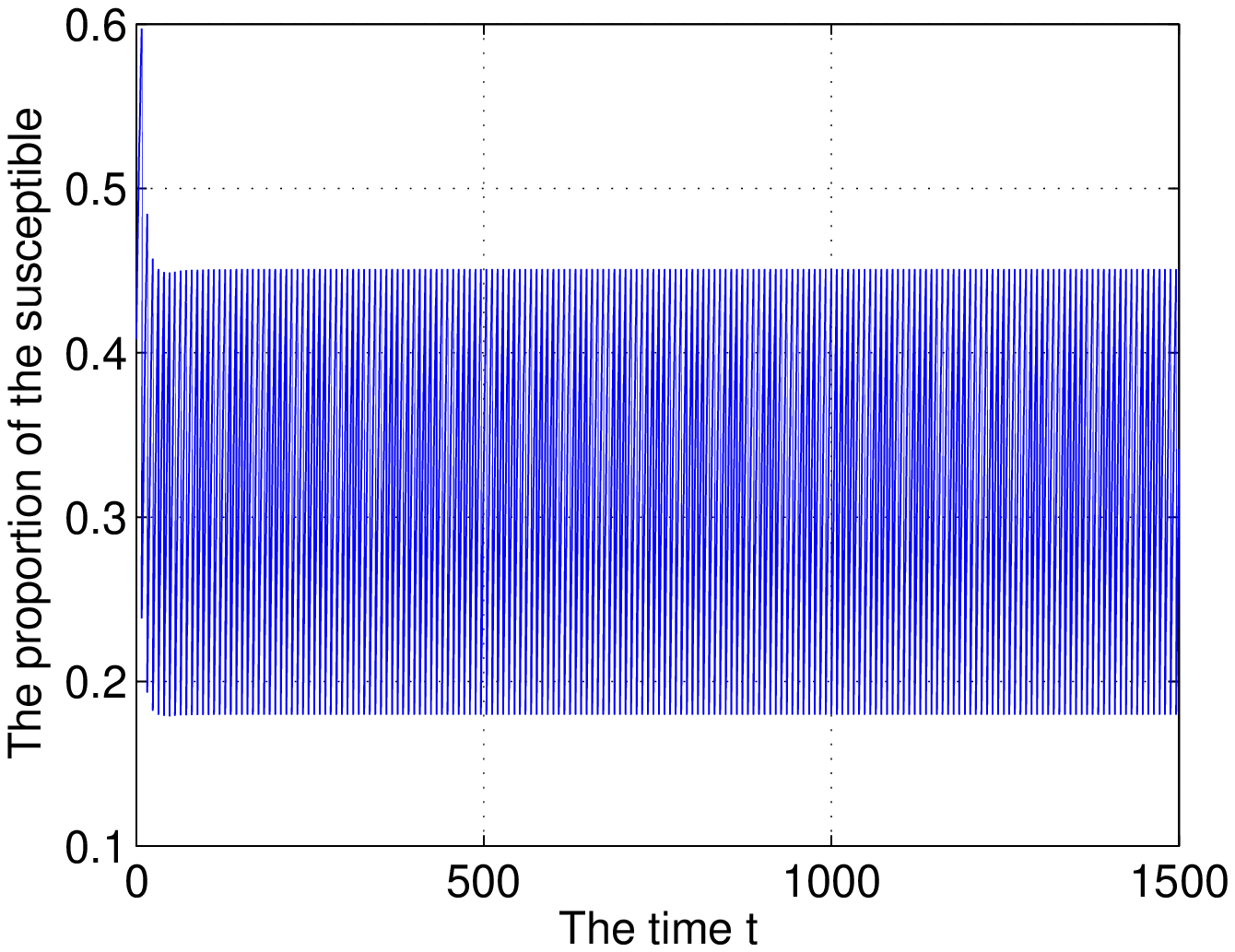}} 
 \subfigure[]{\includegraphics[width=0.33\textwidth]{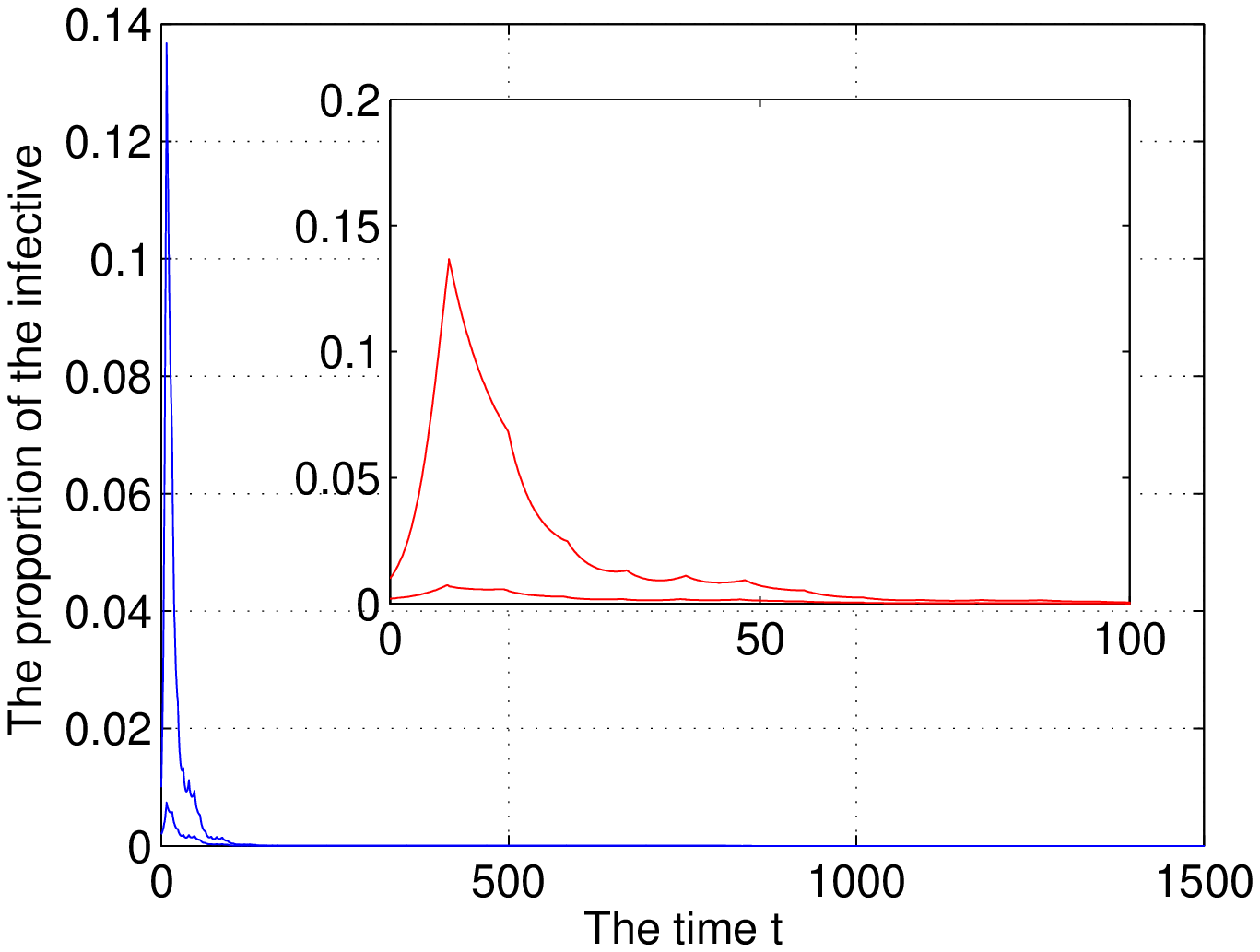}}
 \subfigure[]{\includegraphics[width=0.30\textwidth]{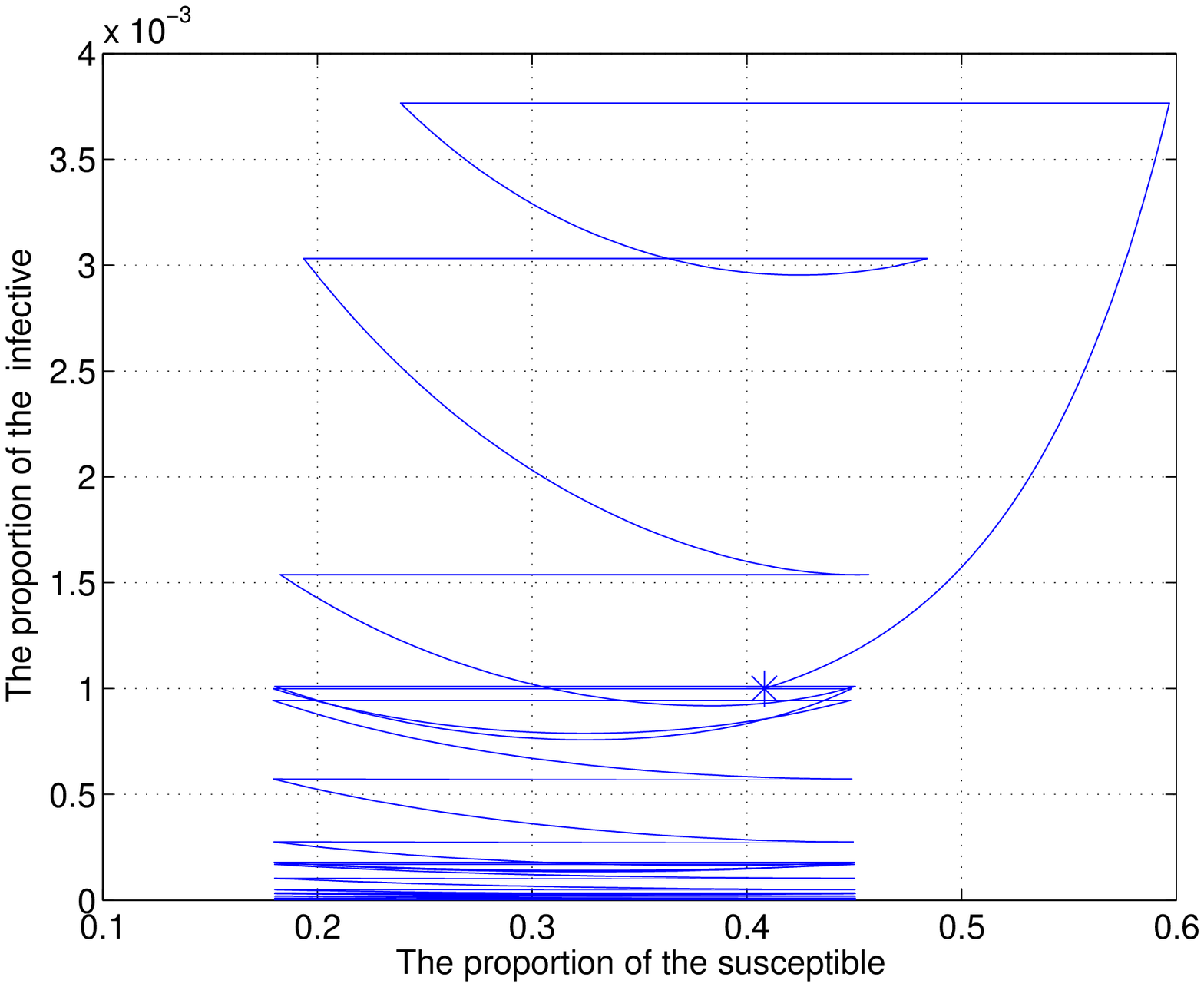}}\\

  \hspace{-1cm}
  \subfigure[]{\includegraphics[width=0.33\textwidth]{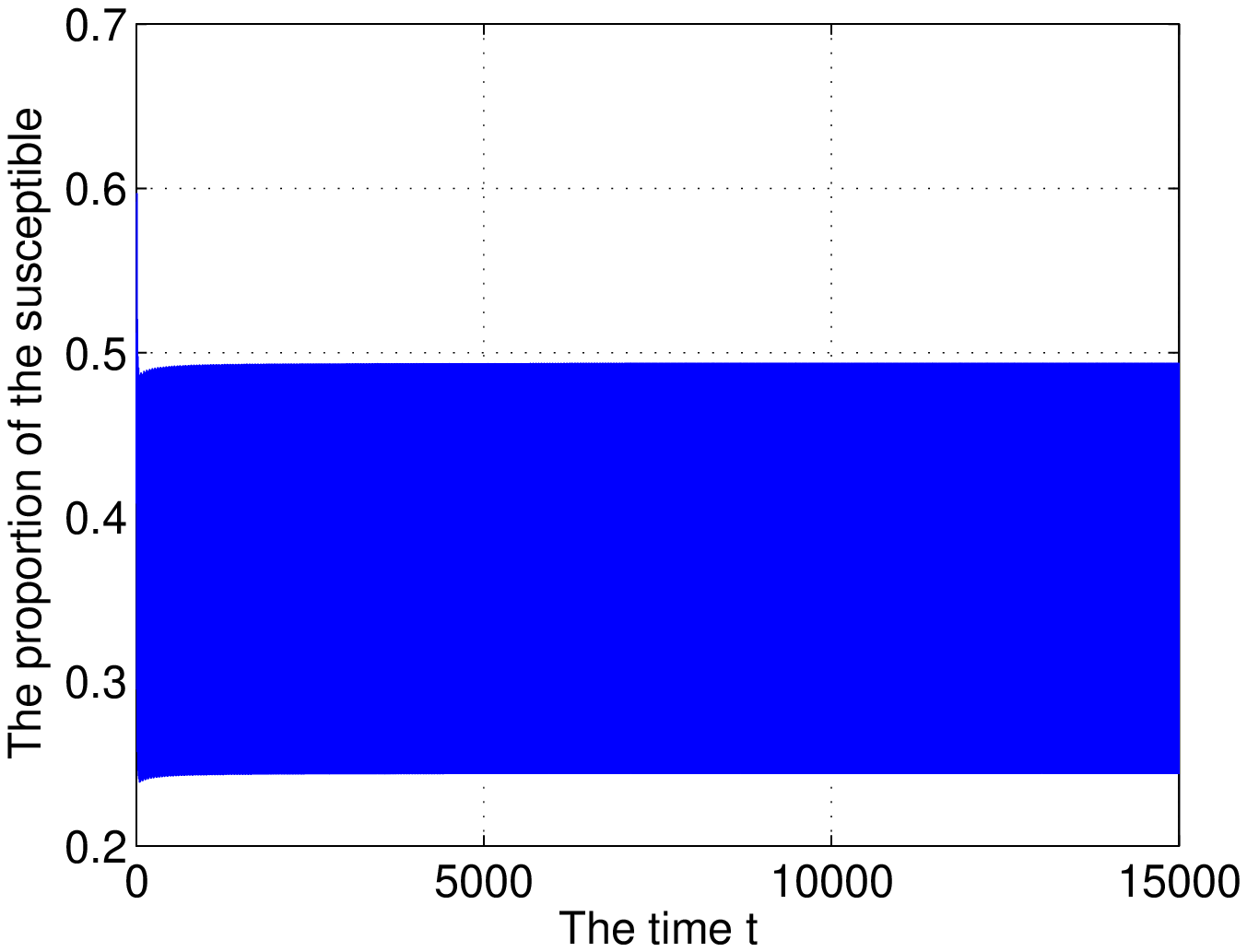}}
  \subfigure[]{\includegraphics[width=0.30\textwidth]{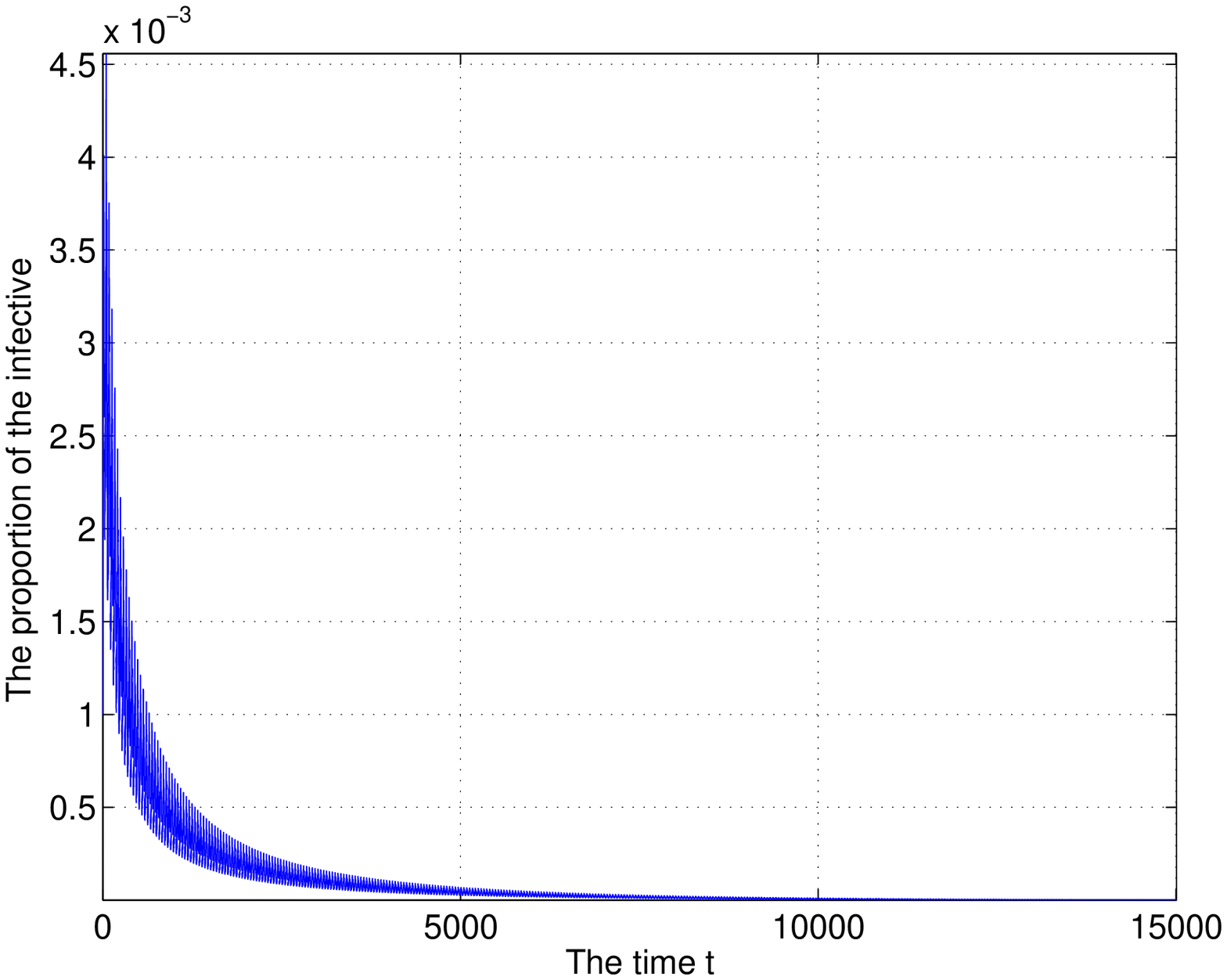}}
  \subfigure[]{\includegraphics[width=0.30\textwidth]{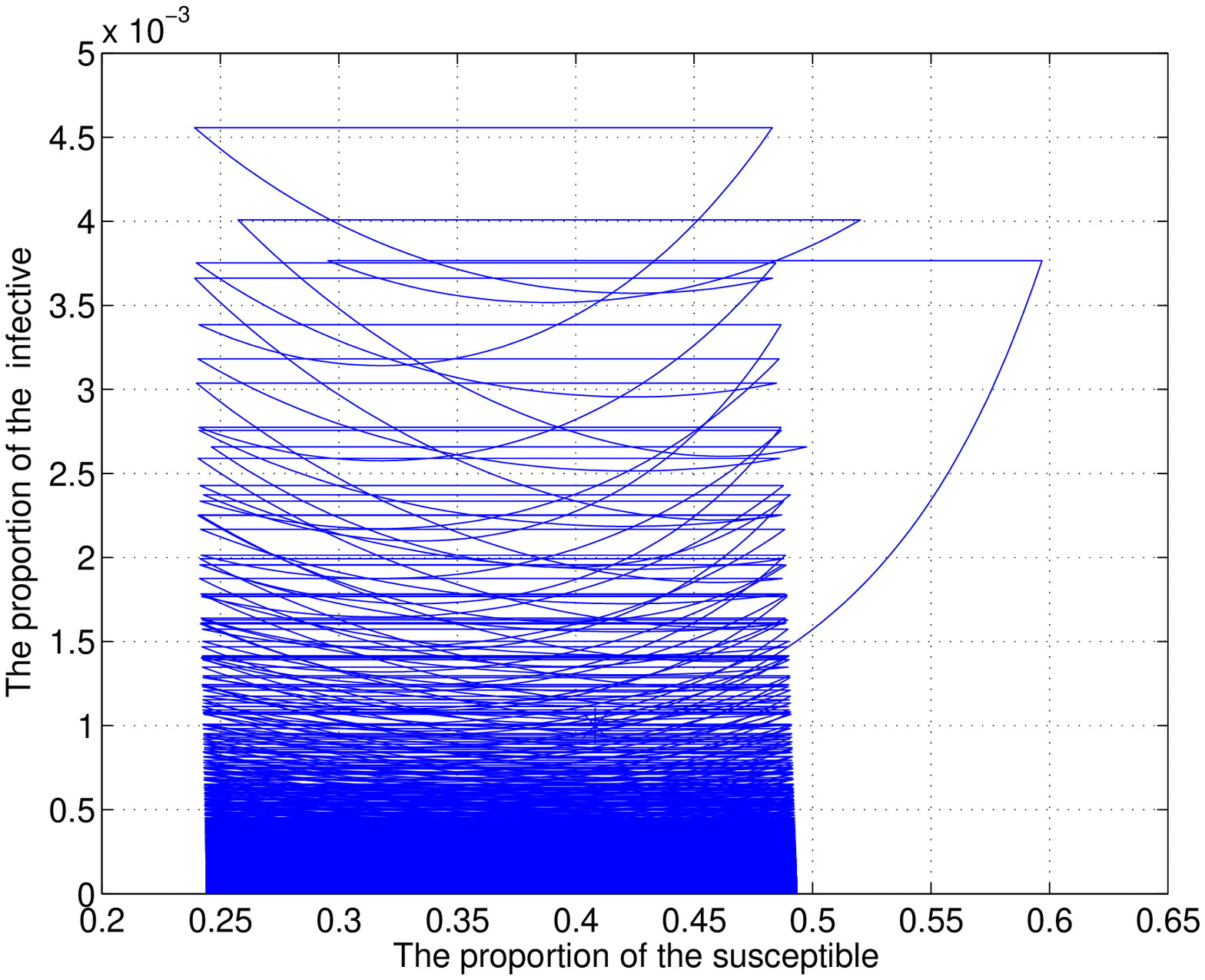}}

 \centering
  \caption{\small Solution of system (2.3) and (2.4) with
  $R_{0}<1$. Subfigures 1(a),(d), and (b), (e) demonstrate respectively the
  susceptible and infective population with respect to time, and Figure
  1 (c), (f) show their corresponding phase-portrait i.e in $S-I$ plane.}\label{fig1}
\end{minipage}

\vspace{4mm} To demonstrate the global stability of the system (2.3)
and (2.4) we take following set parameter values: $b=0.05$, $K=1$,
$\gamma=0.25$, $\alpha=0.002$ and $\beta(t)=0.8[1+0.2\cos(0.05\pi
t)]$. Figure~\ref{fig1} depicts the global stability of system (2.3)
and (2.4) when $R_0<1$ with initial conditions $S(0)=0.408166$,
$I(0)=0.002$, $R(0)=0.08$. Taking a fixed value of $\tau_k=8$ we
have varied  value the parameter $p_k$. Fig.~\ref{fig1}(a)--(c)
shows the solution for $p_k=0.6$, $R_0=0.85909$ and for
Fig.~\ref{fig1}(d)-(f) we take $p_k=0.505$, $R_0=0.99892$
($R_0\rightarrow 1^-$).

\begin{minipage}[c]{1\textwidth}
  \hspace{-1cm}
 \subfigure[]{\includegraphics[width=0.33\textwidth]{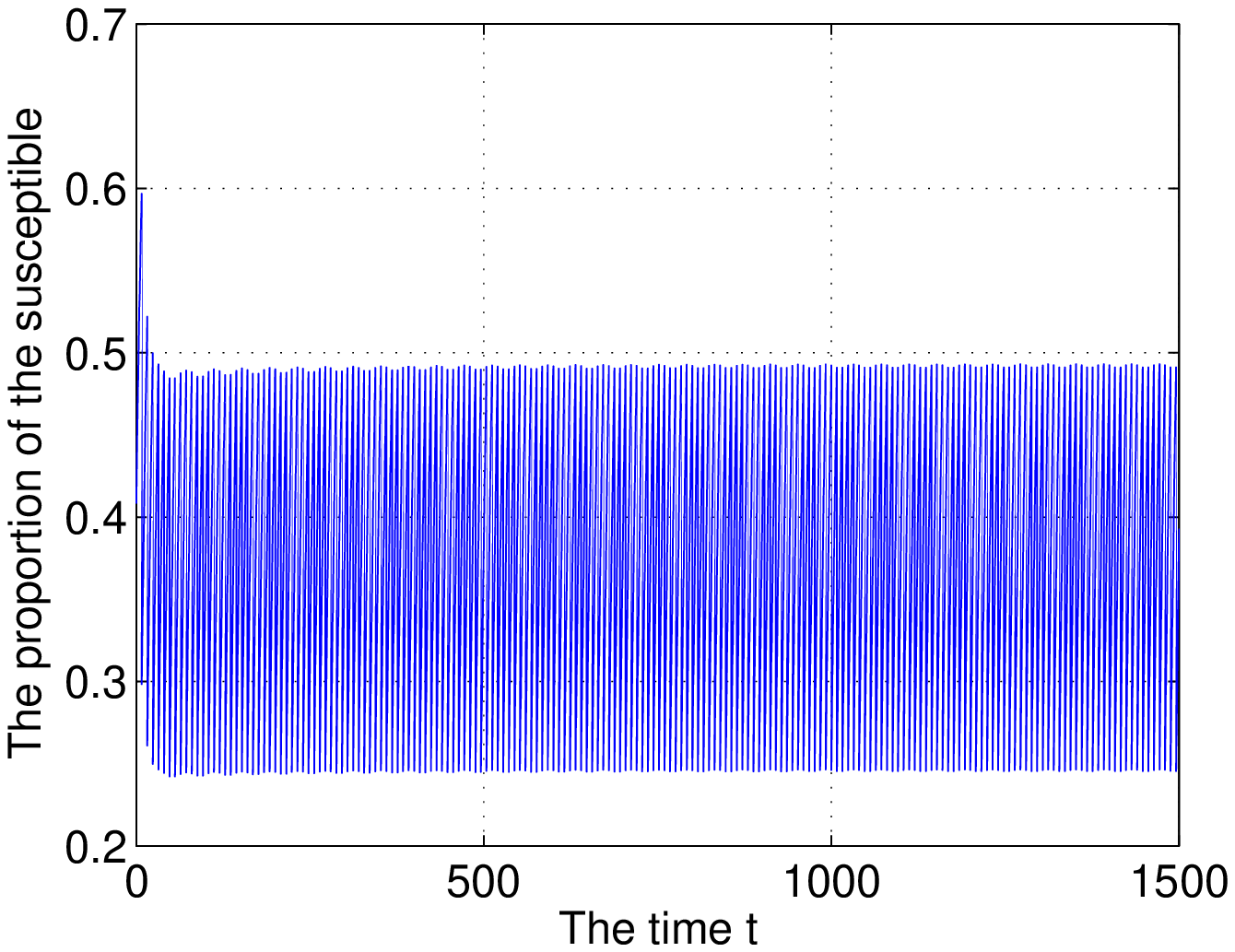}} 
 \subfigure[]{\includegraphics[width=0.33\textwidth]{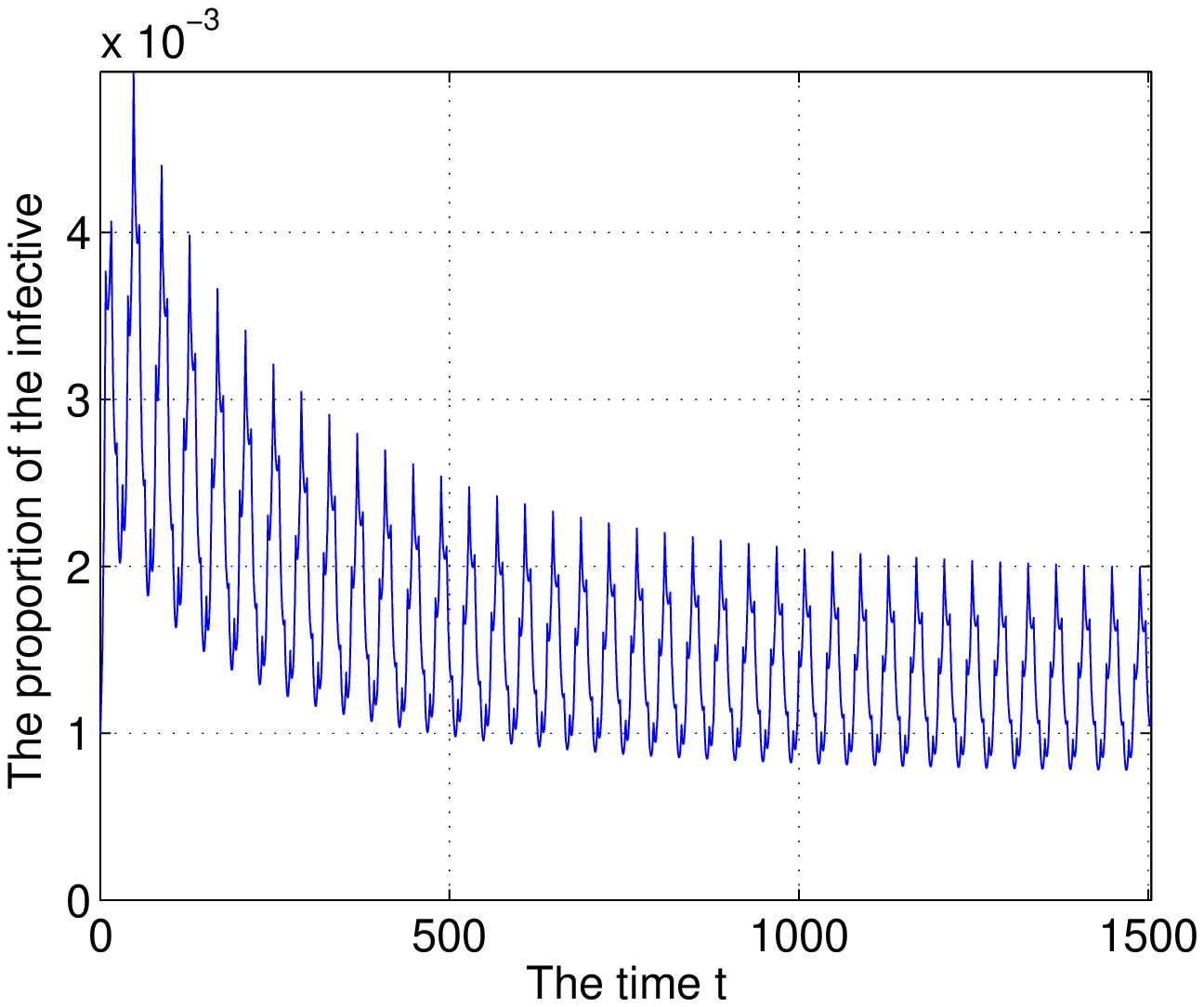}}
 \subfigure[]{\includegraphics[width=0.33\textwidth]{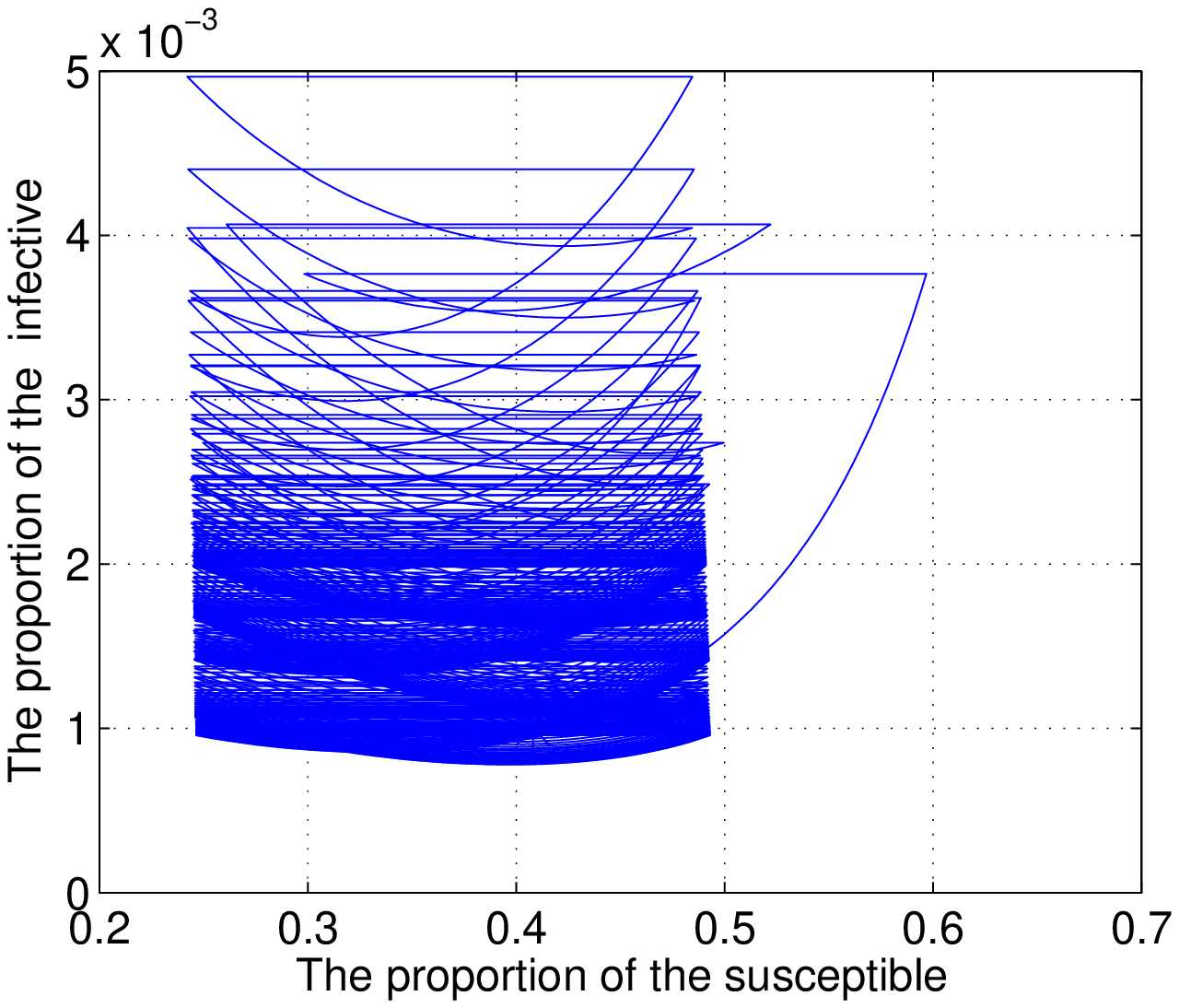}}\\

  \hspace{-1cm}
  \subfigure[]{\includegraphics[width=0.33\textwidth]{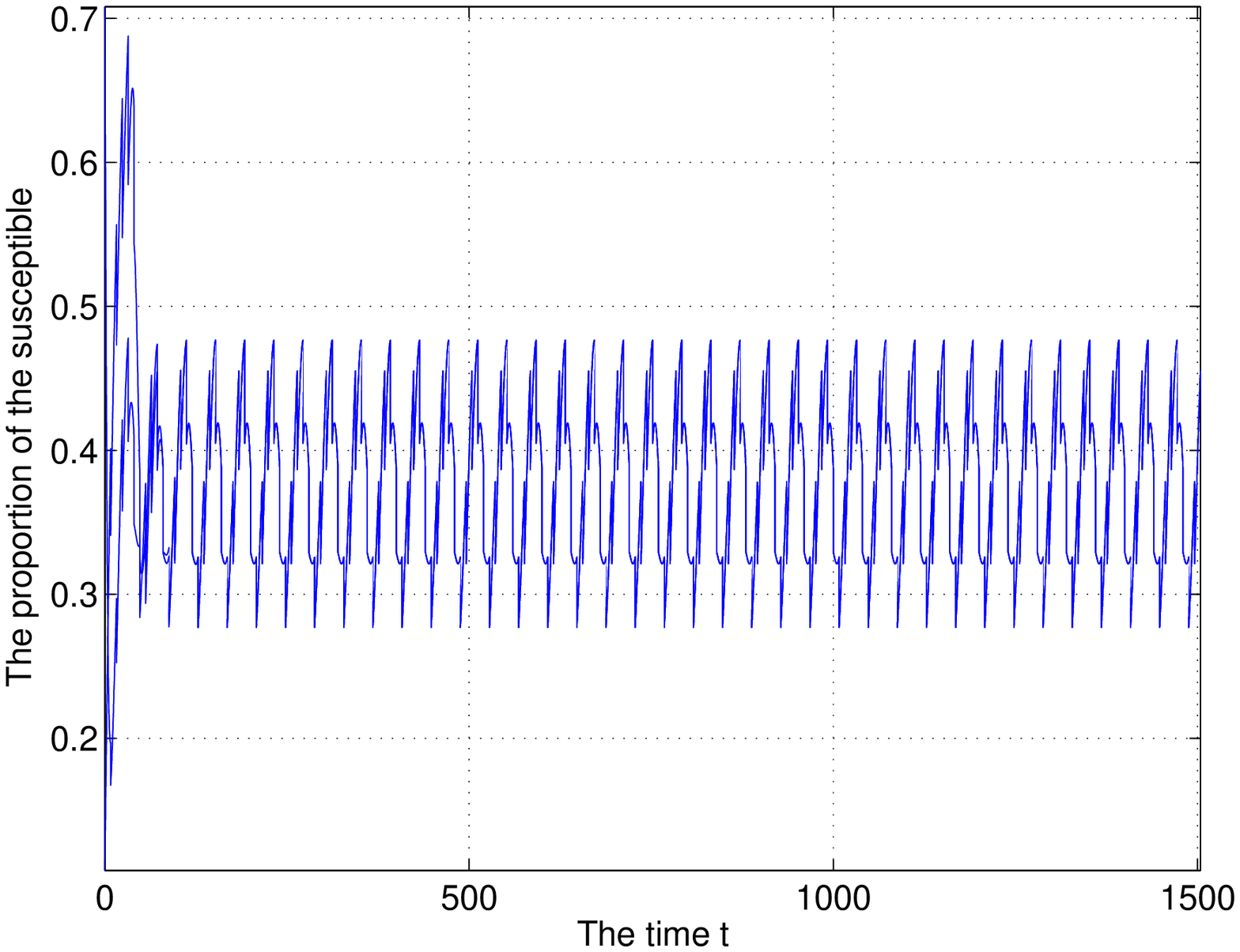}}
  \subfigure[]{\includegraphics[width=0.33\textwidth]{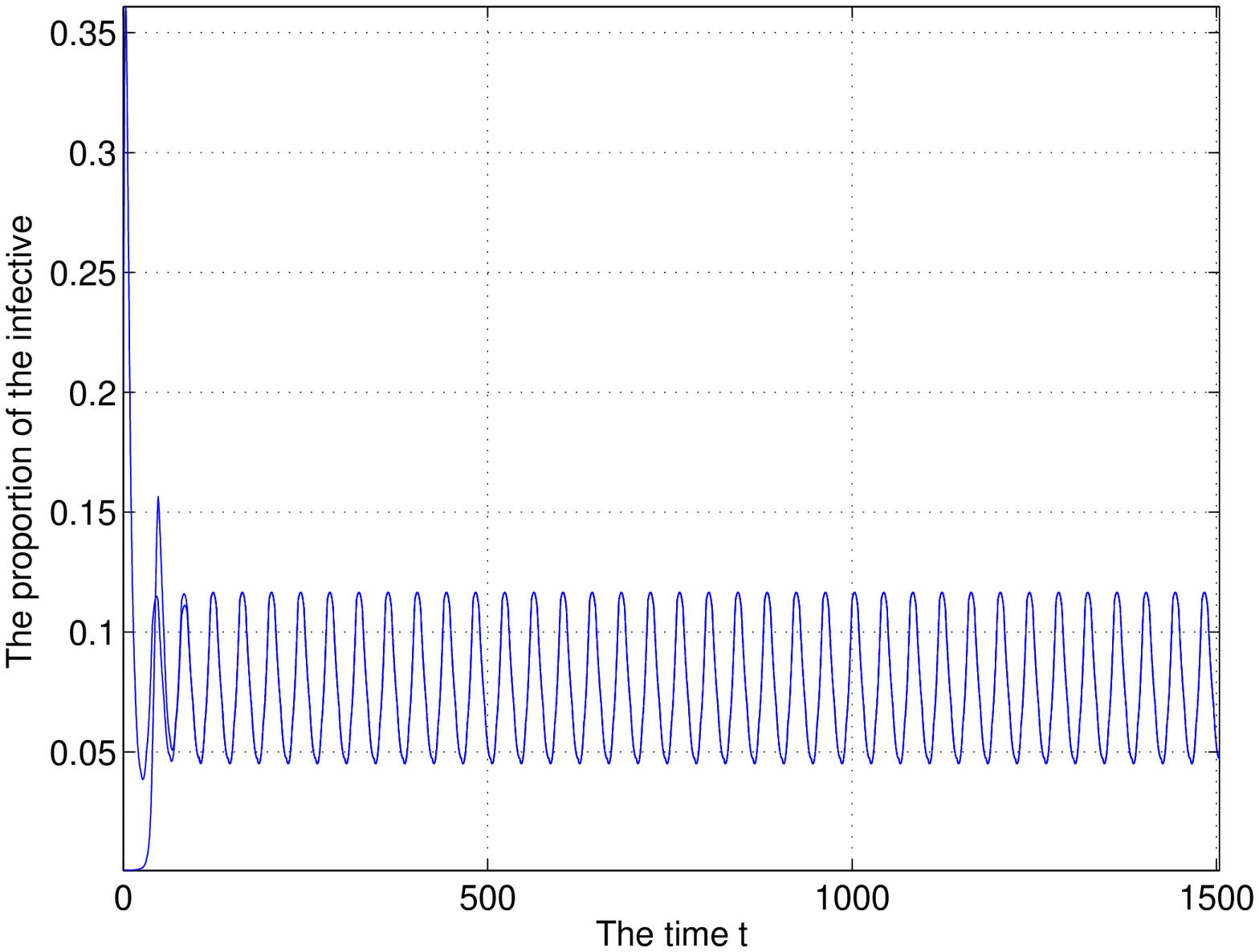}}
  \subfigure[]{\includegraphics[width=0.33\textwidth]{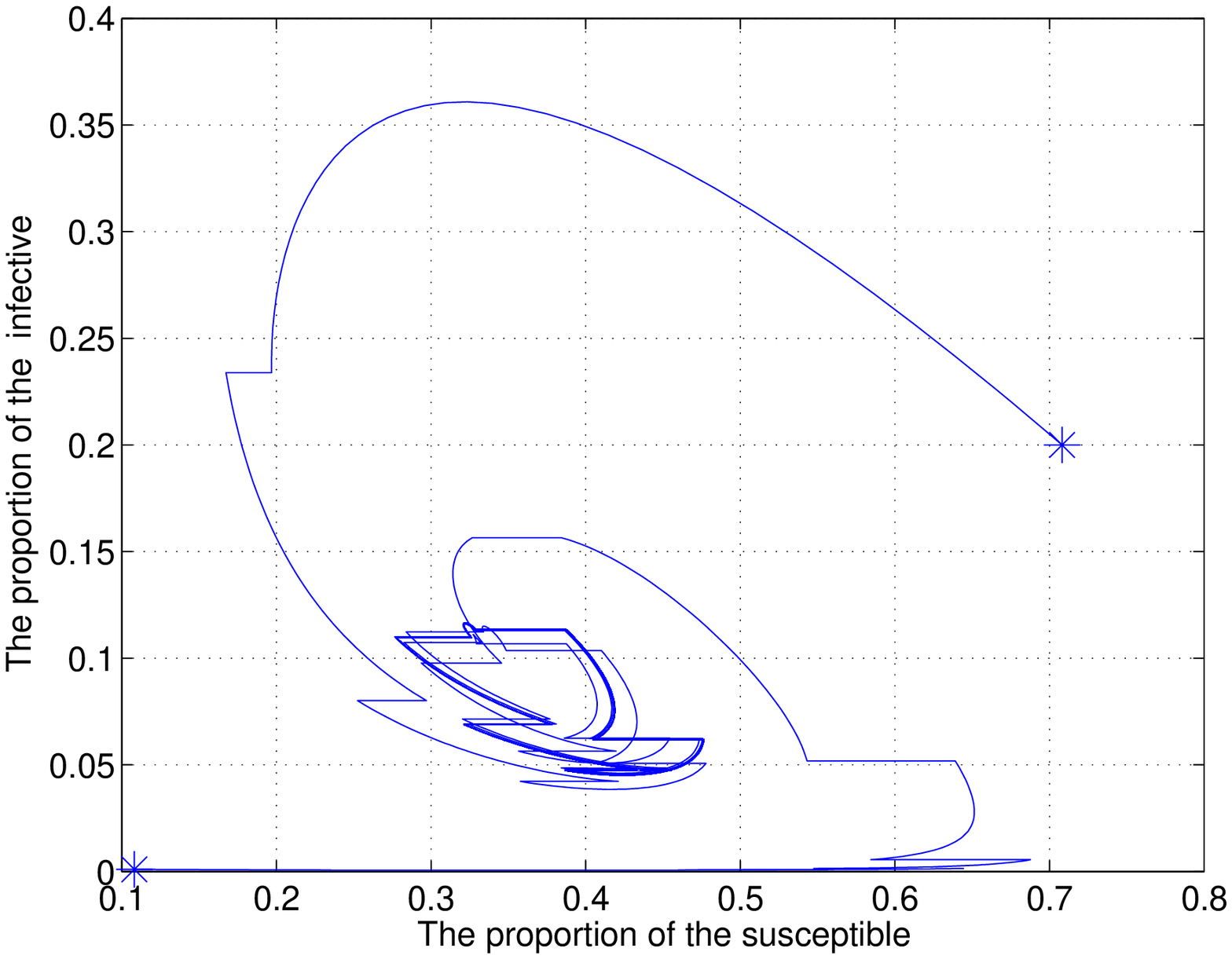}}

  \hspace{-1cm}
  \subfigure[]{\includegraphics[width=0.33\textwidth]{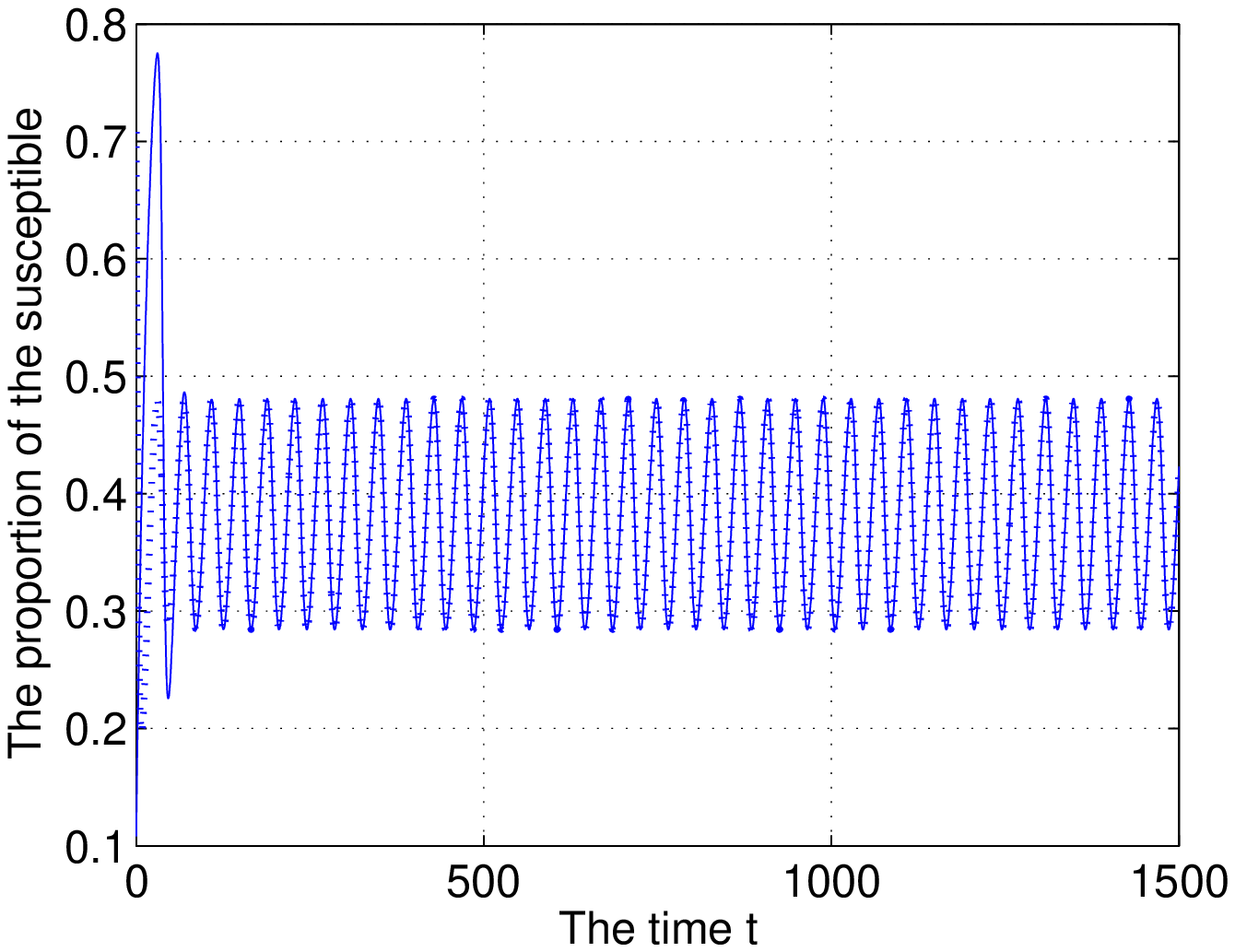}}
  \subfigure[]{\includegraphics[width=0.33\textwidth]{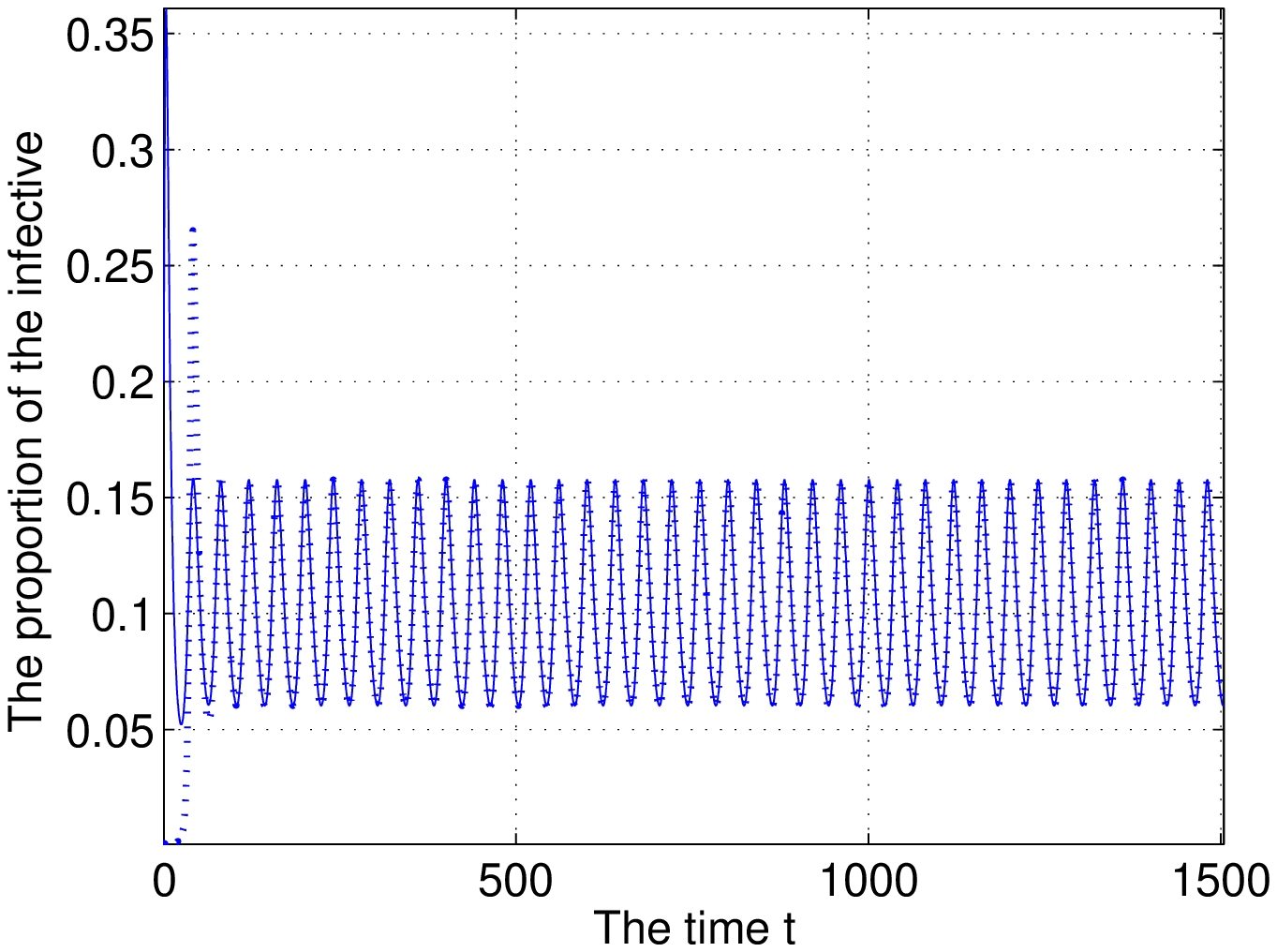}}
  \subfigure[]{\includegraphics[width=0.33\textwidth]{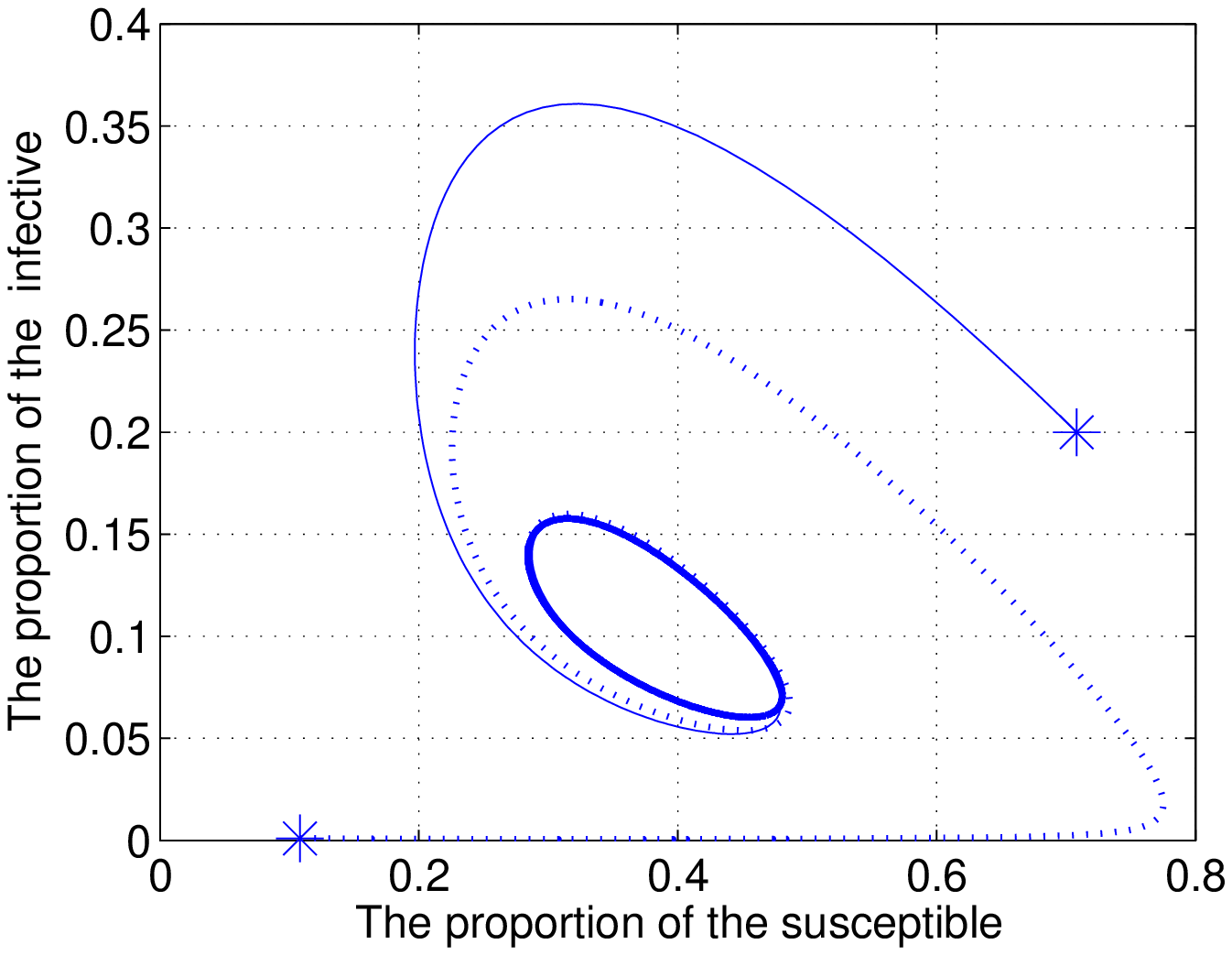}}

  \caption{\small Unstable infection-free situation of the system (2.3) and (2.4)
  at  $R_{0}>1$. Subfigure 2(a),(d),(g) and (b),(e),(h)
  demonstrate respectively the susceptible and infective
  population with respect to time t. Subfigure
  2(c),(f) and (i) show their respective phase portrait i.e in $S-I$ plane.} \label{fig2}
\end{minipage}

\vspace{4mm}

By choosing $R_0>1$ we have shown the unstable infection free
solution  of the system (2.3) and (2.4) in figure~\ref{fig2}. Here
we have chosen two sets of initial conditions: $S(0)=0.108166$,
$I(0)=0.001$, $R(0)=0.08$ and $S(0)=0.708166$, $I(0)=0.2$,
$R(0)=0.08$. The Fig.~\ref{fig2}(a)--(c) with $p_k=0.5$ and
$R_0=1.00702$; The Fig.~\ref{fig2}(d)--(f) with $p_k=0.15$ and
$R_0=1.88779$; The Fig.~\ref{fig2}(g)--(i) is the limiting case with
$p_k=0$ and $R_0=2.64901$. The other parameters have the same value
as in Fig.~\ref{fig1}.

\vspace{4mm}

Now we fix the parameter $p_k$ and vary the parameter $\tau_k$. In
figure~\ref{fig3}, the initial values are used $S(0)=0.408166$,
$I(0)=0.002$, $R(0)=0.08$. Fig.~\ref{fig3}(a)--(c) the parameters
are chosen as $p=0.5$ and $\tau=5$ with $R_0=0.72972$;
Fig.~\ref{fig3}(d)--(f) the parameters are chosen as $p=0.5$ and
$\tau_k=13$ with $R_0=1.34774$, the other parameters as the same
Fig.\ref{fig1}. Fig.~\ref{fig3} suggests the positive periodic
solution is globally asymptotically stable when the $R_0>1$.

\vspace{4mm}

\begin{minipage}[c]{1\textwidth}
  \hspace{-1cm}
 \subfigure[]{\includegraphics[width=0.33\textwidth]{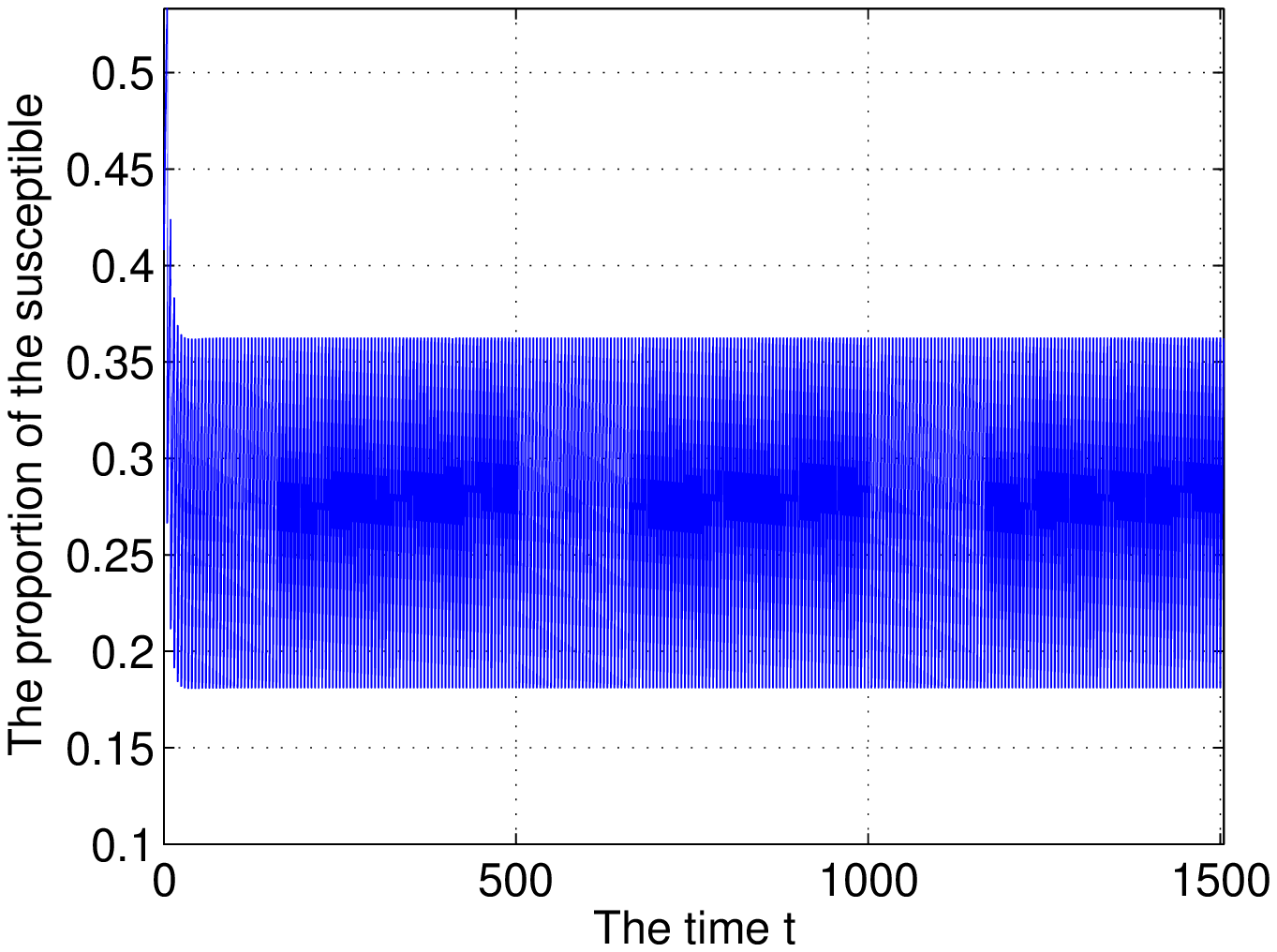}} 
 \subfigure[]{\includegraphics[width=0.33\textwidth]{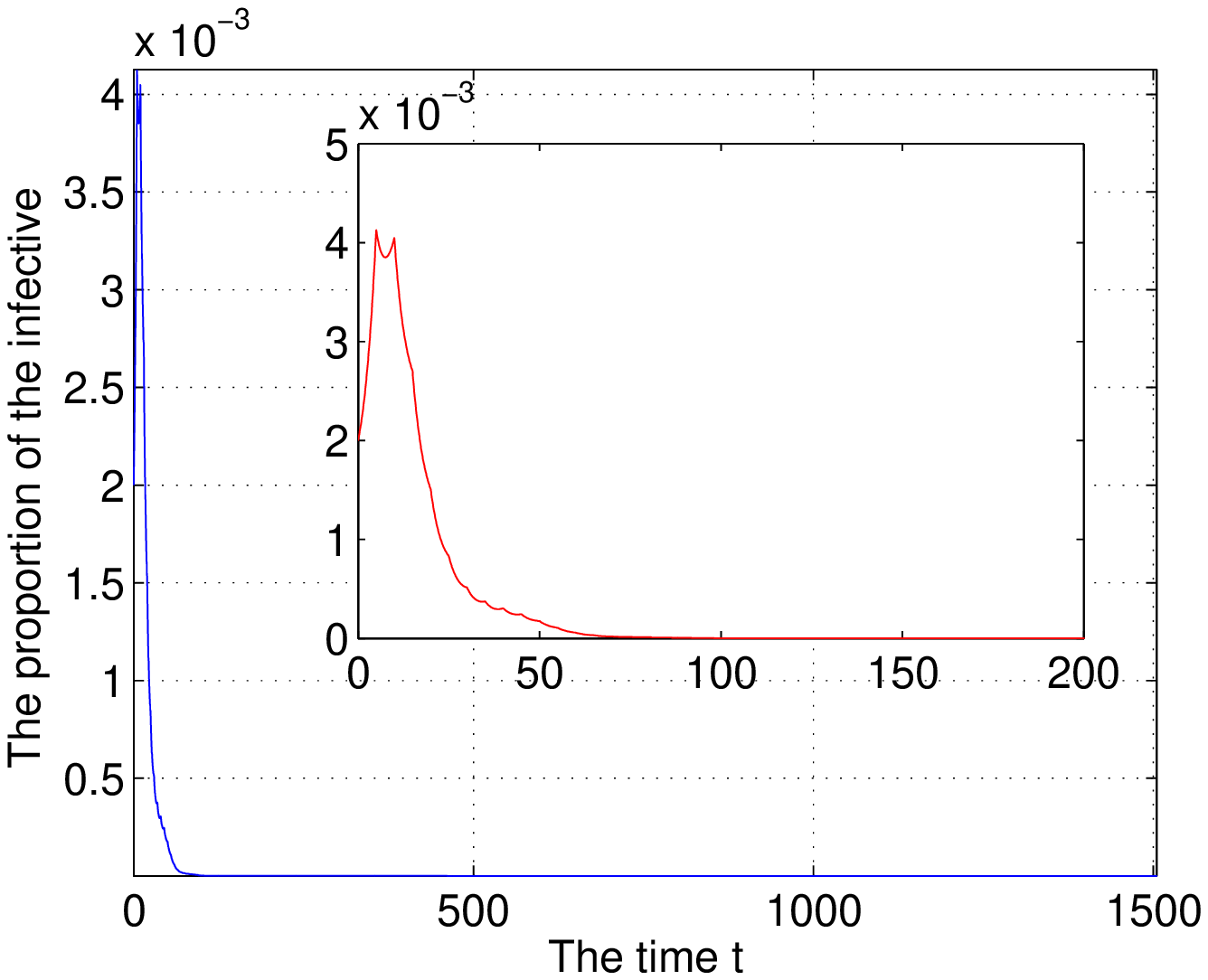}}
 \subfigure[]{\includegraphics[width=0.33\textwidth]{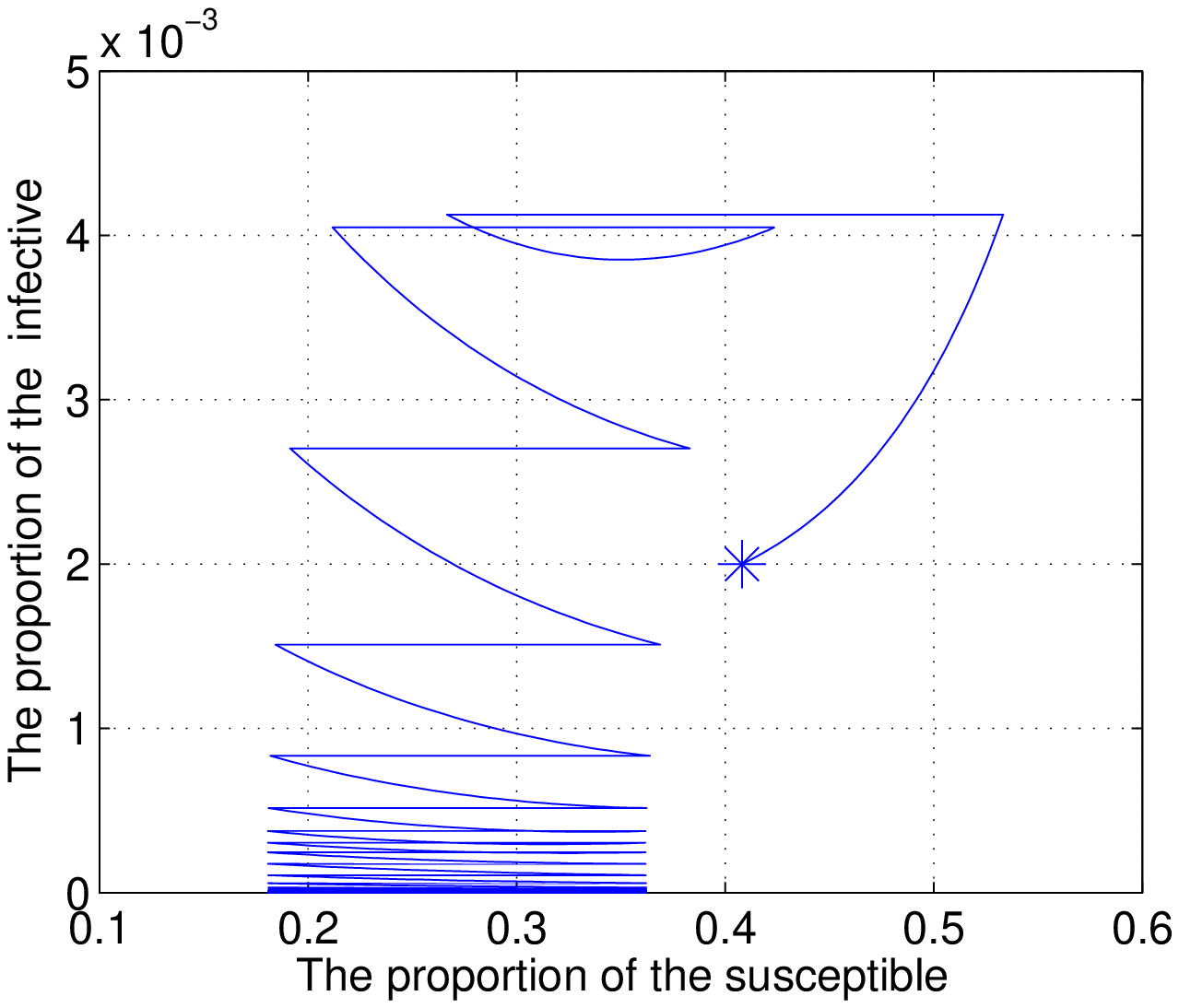}}\\

  \hspace{-1cm}
  \subfigure[]{\includegraphics[width=0.33\textwidth]{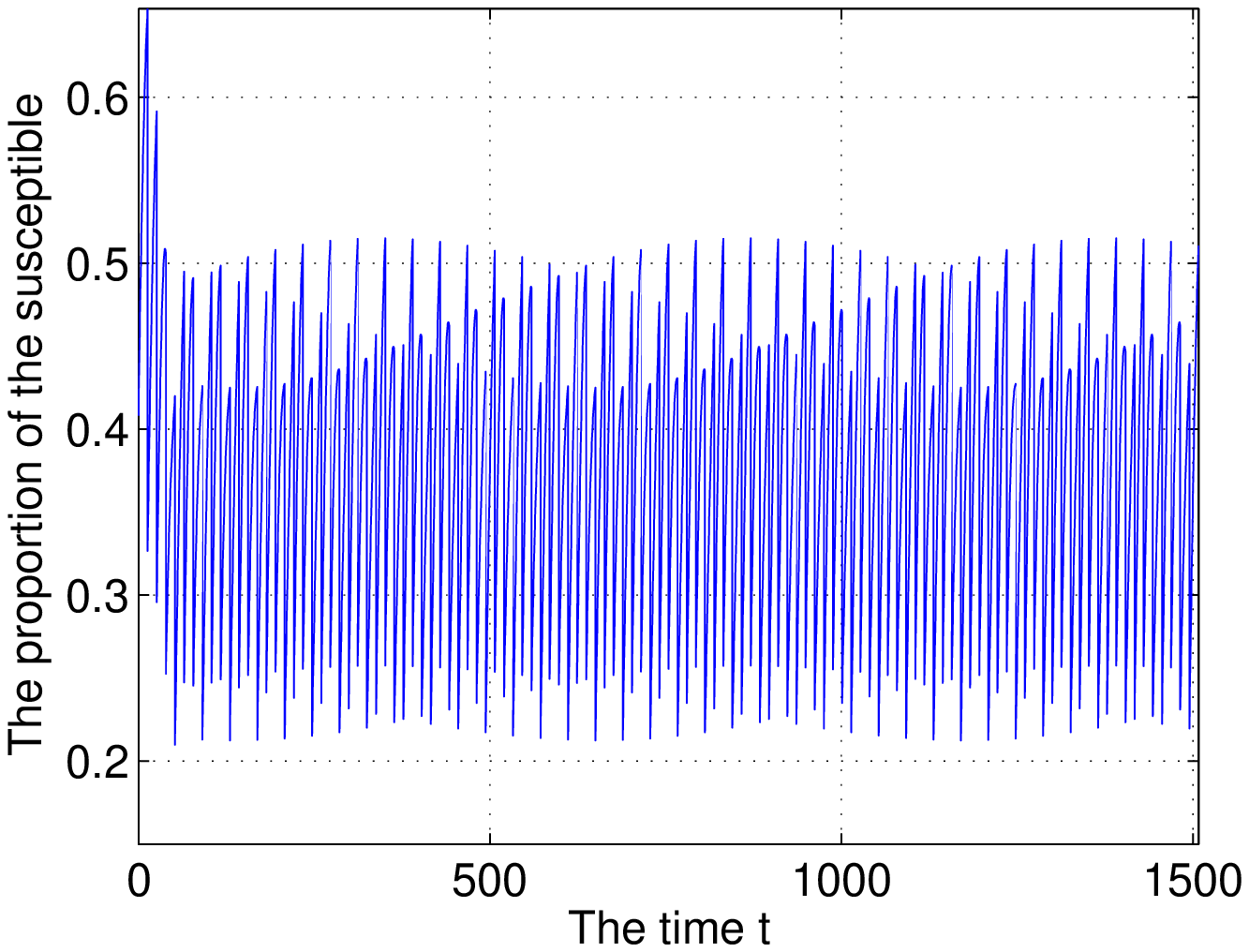}}
  \subfigure[]{\includegraphics[width=0.33\textwidth]{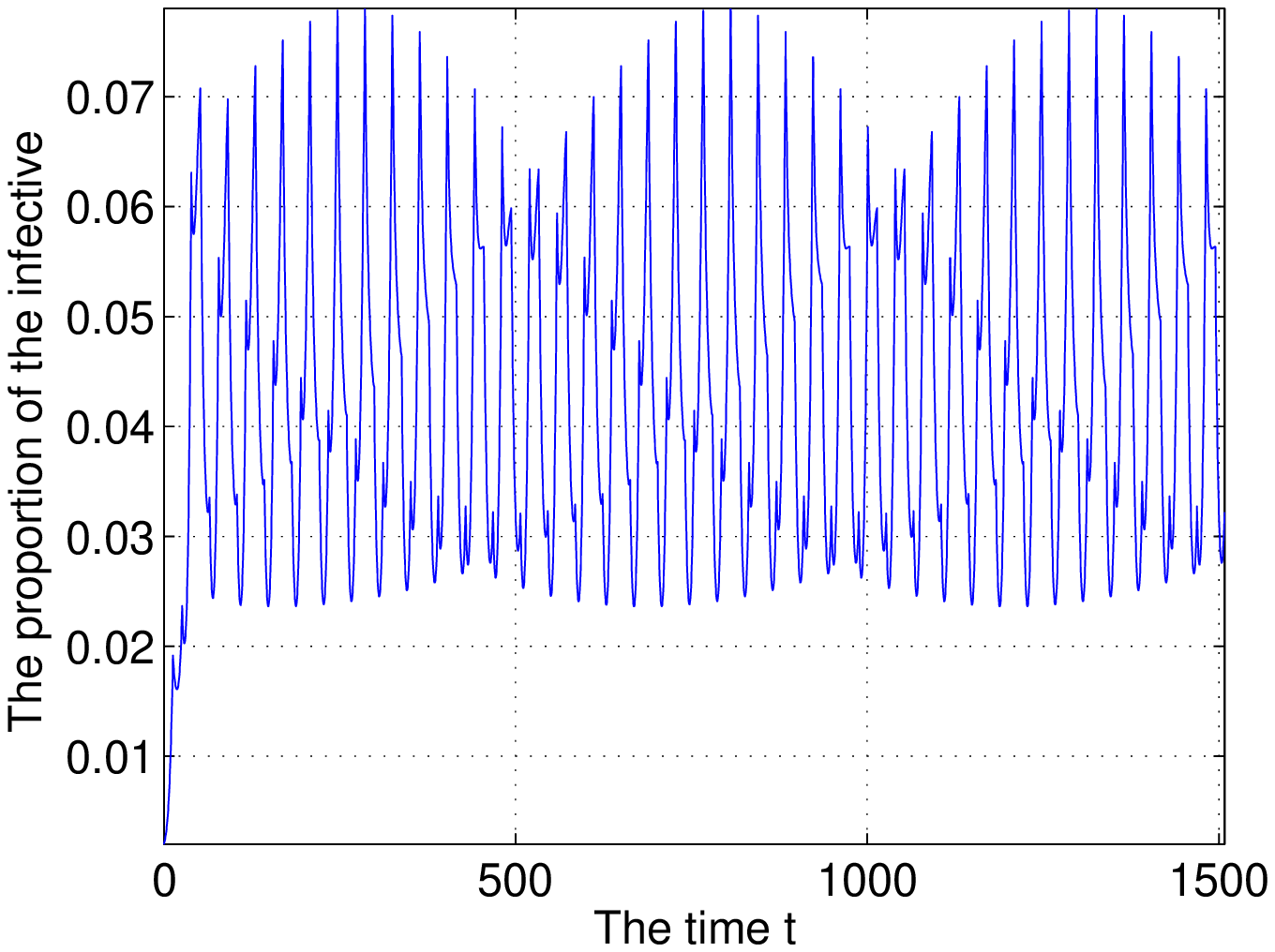}}
  \subfigure[]{\includegraphics[width=0.33\textwidth]{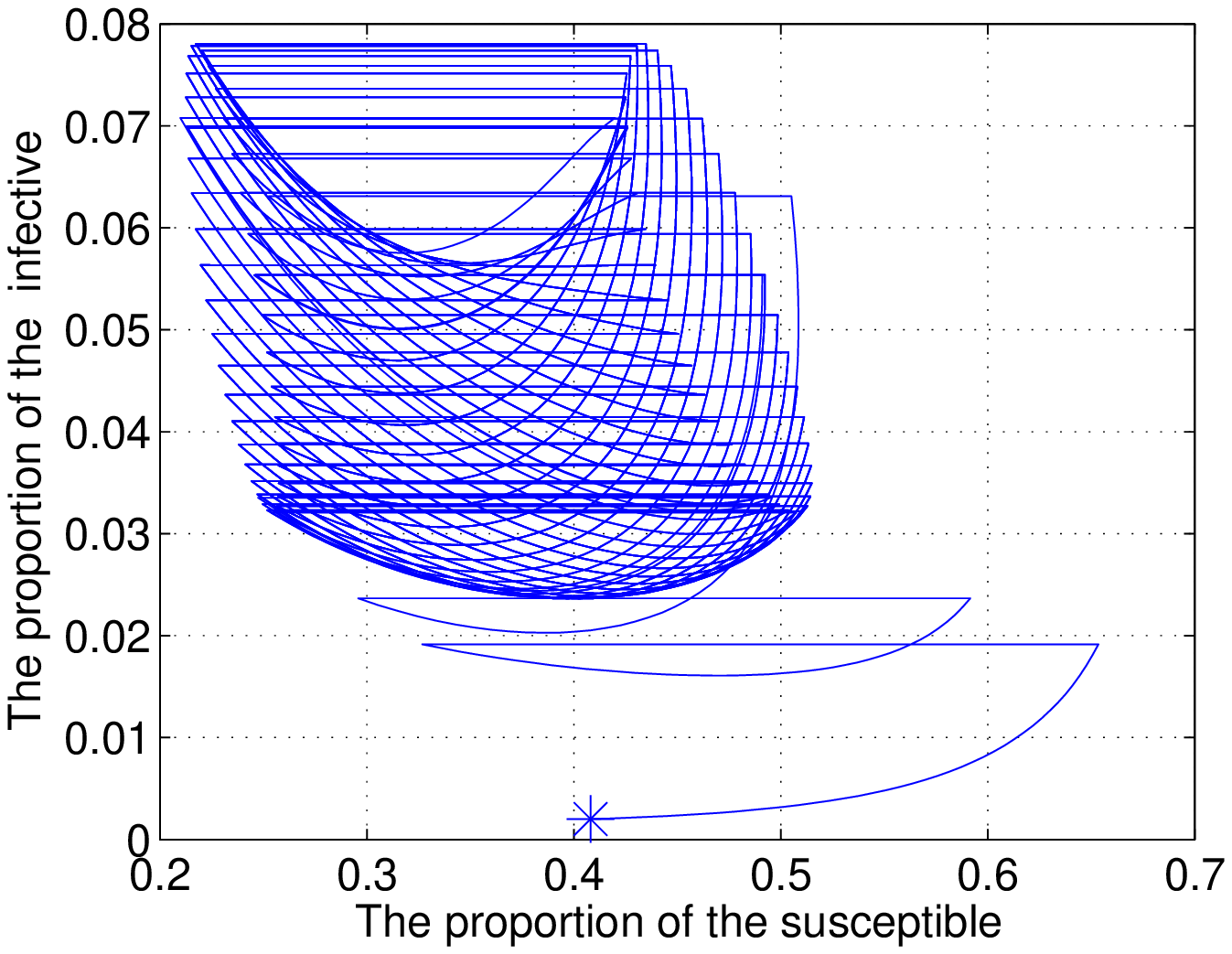}}

 \centering
  \caption{\small The orbits of the system (2.3) and (2.4)with respect to time series.
   Subfigure 3(c), (f) depict the orbits on $S-I$ plane, and (a), (d) and (b), (e) show the
  time series for susceptible and infective respectively.}\label{fig3}
\end{minipage}

\vspace{3mm} {\section *{7. \   Discussion  }}

Vaccination strategies are designed and applied to control or
eradicate an infection from the population. As for many directly
transmitted infectious diseases levels of infection oscillate
regularly it may be more useful to use a time dependent contact rate
in place of a constant contact rate.

The idea of this vaccination strategy is to give the whole
susceptible population vaccine at a periodically varying rate
$\beta(t)$: Many vaccination strategies are applied in different
parts of the world. Some of these use a periodic vaccination
strategy. The sort of periodic vaccination rate most commonly used
is the pulse periodic function [24]. As the study of periodicity and
other oscillatory behavior in the incidence of infectious diseases,
especially childhood diseases, is one of our main goals in this
work, we shall study the effect of a general time dependent periodic
vaccination strategy on the dynamics of these infectious diseases.
This new vaccination strategy is designed to vaccinate susceptible
of all ages using a varying periodic vaccination rate $\beta(t)$.
This periodic vaccination is intended to keep the disease free
solution stable by controlling the number of susceptible over the
vaccination interval. We have study an SIR epidemic model with
seasonality in the transmission rate but the results apply equally
well to the case where the contact rate is constant.

A lot of infectious diseases have, at least, temporary varying
transmission rate due to the seasonal fluctuation or changing social
behavior. The time dependent epidemic models with pulse vaccination
are natural generalization and have more applications in real world.
The study of the dynamical behaviors of epidemic models with time
dependent infection rate and pulse vaccination is an important and
challenging issue for us. The SIR epidemic model with periodic
infection rate and pulse vaccination is one of the simple and
important epidemic models. In this article, we have investigated the
dynamical behavior of a classical periodic SIR model with pulse
vaccination. We have shown that the infection--free periodic
solution $(S^*(t),0,R^*(t))$ is globally stable if the basic
reproductive number $R_0$ is less than 1, and the infection--free
periodic solution $(S^*(t),0,R^*(t))$ is unstable if the basic
reproductive number $R_0$ is greater than 1. In the later case, the
disease will uniform persist in the population. Moreover, we have
used a standard bifurcation technique to show the existence of the
positive periodic solutions which arise near the infection--free
periodic solution $(S^*(t),0,R^*(t))$.

The basic reproductive number is a fundamental parameter and one of
the most useful quantities characterize the magnitude of an
infectious disease transmission [23]. It is the average number of
secondary cases that arise from a primary case in a fully
susceptible population. For the classical SIR model with constant
infection rate, the basic reproductive number $R_0$ is the product
of the average infection period and the average infection in unit
time. From a deterministic viewpoint, the eradication of an
infection requires $R_0<1$, and the condition for outbreak or for
maintenance of epidemic infection demands $R_0>1$. The basic
reproductive number $R_0$ for the SIR model with periodic infection
rate and pulse vaccination is defined using the average of
$\beta(t)S^*(t)$ over one period. It is proved that the disease will
die out if $R_0 < 1$, and the disease will persist when $R_0
> 1$. The classical SIR model has a globally stable endemic
equilibrium. The periodic solution of the SIR model with periodic
infection rate and pulse vaccination is also expected to play the
similar role. However, here we have got comparatively few new
results on the existence of the positive periodic solution of the
epidemic models. The bifurcation theorem has been used to establish
the existence of periodic solution and found that positive periodic
solution of our system is globally stable when $R_0> 1$. On the
other hand, infection free periodic solution is globally stable if
$R_0< 1$. We have simulated the global behaviors of the model for
both cases. From the numerical simulation, we conclude that the
positive periodic solution of the SIR model is unique and globally
stable. The uniqueness and stability of the periodic solution is of
course more challenging problem for the researcher [24]. We expect
the uniqueness and global stability of the periodic solution will be
proved very soon.


\vspace{1cm}

\begin{thebibliography}{99}
\bibitem{1}H. W. Hethcote, The Mathematics of Infectious Diseases, SIAM Review,
42:4(2000), 599-653.
\bibitem{2} H. W. Hethcote, Three basic epidemiological models, in Applied Mathematical
Ecology, L. Gross, T. G. Hallam, and S. A. Levin, eds.,
Springer-Verlag, Berlin, 1989, 119-144.
\bibitem{3} L.Stone, B.Shulgin and Z.Agur, Theoretical examination of the pulse vaccination policy
in the SIR epidemic modelc, Mathl. Comput. Modelling 31 (4/5),
207-215 (2000).
\bibitem{4}  D.J. Nokes, J. Swinton, The control of childhood viral
infections by pulse vaccination, IMA Journal of Mathematics Applied
in Medicine d Biology 12, 29-53 (1995).
\bibitem{5}A. d¡¯onofrio, Pulse vaccination strategy in the SIR epidemic
model: Global asymptotic stable eradication in presence of vaccine
failures, Mathl. Comput. Modelling 96 (4/5), 473-489 (2002).
 \bibitem{6}A. d¡¯onofrio, Stability properties of pulse vaccination
strategy in SEIR epidemic model, Mathematical Biosciences 179 (l),
57-72 (2002).
\bibitem{7}A. d'Onofrio,Globally stable vaccine-induced eradication
of horizontally and vertically transmitted infectious diseases with
periodic contact rates and disease-dependent demographic factors in
the population, Applied Mathematics and Computation 140,
537-547(2003).
\bibitem{8} Yicang Zhou, Hanwu Liu, Stability of Periodic Solutions for an SIS Model with Pulse
Vaccination, Mathematical and Computer Modelling, 38,299-308 (2003).
\bibitem{9} A. d'Onofrio, Mixed pulse vaccination strategy in epidemic model with
realistically distributed infectious and latent times, Applied
Mathematics and Computation 151, 181-187(2004).
\bibitem{10} K. M. Fuhrman, I. G. Lauko and G. A. Pinter, Asymptotic Behavior of an SI Epidemic
Model with Pulse Removal, Mathl. Comput. Modelling 40,
371-386(2004).
\bibitem{11}Z. Jin, The study for ecological and epidemical models influenced by impulses, Doctoral Thesis, Xian Jiaotong
University, (2001).
\bibitem{12}Z.L. Agur, J.L. Deneubourg, The effect of
environmental disturbance on the dynamics of marine intertidal
populations, Theor. Pop. Biol. 27, 75-90 (1985).
\bibitem{13} Z.L. Agur etal., Pulse mass measles vaccination across age cohorts, Proc.
National Acad Scz, USA. 90, 11698 11702 (1993).
\bibitem{14}  E. Pourabbas, A. d'Onofrio, M. Rafanelli, A method to estimate the incidence of
communicable diseases under seasonal fluctuations with application
to cholera, Appl. Math. Comput. 118,161-174(2001).
\bibitem{15}  H.L.Smith, Subharmonic bifurcation in a SIR epidemic model, J.
Math. Biol. 17, 163-177(1983) .
\bibitem{16}  I.B. Schwartz, Small
amplitude, long period outbreaks in seasonally driven epidemic, J.
Math. Biol. 30,  473-491(1992).
\bibitem{17} Rabinowitz, P. H., Some global results for nonlinear eigenvalue problems. J. Functional Analysis,
7, 487-513 (1971).
\bibitem{18} V. Lakshmikantham, D.D.Bainov and  P.S.Simeonov,
Theory of impulsive differential equations, World Scientific, 1989.
\bibitem{19}  Drumi Bainov, Pavel Simeonov, Impulsive differential
equations: Periodic solutions and applications, Longman Scientific
and Technical , 1993.
\bibitem{20}Bing Liu, Lansun Chen, The periodic competing
Lotka-Volterra model with impulsive effect, Mathematical Medicine
and Biology, 21, 129-145 (2004).
\bibitem{21}Jing Hui, Lansun Chen, Existence of Positive Periodic Solution of Periodic Time-Dependent
Predator-Prey System with Impulsive effects, 20(3),423-432(2004).
\bibitem{22}Vainberg, M.V., Trenogrn, v.A., The methods of Lyapunov
and Schmidt in the theorey of nonlinear equations and further
development. Russian Math. Surveys, 17(2), 1-60(1962).
\bibitem{23}B. G. Williams, C. Dye, Infectious disease persistence when transmision varies
seasonally, Mathematical Biosciences, 157(1997), 77-88.
\bibitem{24}Boris Shulgin, Lewi Stone, Zvia Agur, Pulse Vaccination Strategy in the SIR
Epidemic Model, Bulletin of Mathematical Biology, (1998) 60,
1123-1148.
\bibitem{25} Zhen Jin, Ma zhien, Periodic solutions for delay differential equations model of
plankton allelopathy,  Computers and Math $\&$ Appl. 44(2002)
491-500.
\bibitem{26}Zhen Jin, Ma Zhien, Han Maoan, The existence of periodic solutions of the n-species
Lotka-Volterra competition systems with impulsive, Chaos, Solitons
and Fractals, 22(2004) 181-188.

\end{thebibliography}
\end{document}